\documentclass[12pt]{article}
\usepackage[totalwidth=480pt,totalheight=640pt]{geometry}
\usepackage[centertags]{amsmath}
\usepackage{array,multirow}
\usepackage{amssymb}
\usepackage{graphicx}
\usepackage{color}
\usepackage{setspace}
\pdfoutput=1

\usepackage{epsfig}

\def\Or[#1]{{\text{O}}\left({#1}\right)}
\def\dotl[#1,#2]{\left\langle #1, #2 \right\rangle}
\def\dotlb[#1,#2]{[ #1, #2 ]}
\def\dotp[#1,#2]{(#1) \cdot (#2)}
\def\aff[#1,#2]{\hat{#1}(#2)}
\def\n4sym{{\cal N}=4 SYM}
\def\>{\rangle}
\def\<{\langle}
\def\weight[#1,#2,#3]{\{(#1),#2,#3\}}
\def\ads[#1]{$\text{AdS}_{#1}$}

\newcommand{\be}{\begin{equation}}
\newcommand{\ee}{\end{equation}}
\newcommand{\ba}{\begin{eqnarray}}
\newcommand{\ea}{\end{eqnarray}}
\newcommand{\bea}{\begin{eqnarray}}
\newcommand{\eea}{\end{eqnarray}}

\newcommand{\CC}{{\cal C}}

\newcommand{\NN}{{\cal N}}
\newcommand{\CO}{{\cal O}}

\newcommand{\nn}{\nonumber}

\usepackage{hyperref}
\title{Mellin Amplitudes}

\begin{document}

\begin{titlepage}

\begin{center}
\vspace{1cm}

{\Large \bf  Analyticity and the Holographic S-Matrix}

\vspace{0.8cm}

\bf{A. Liam Fitzpatrick$^1$, Jared Kaplan$^2$}

\vspace{.5cm}

{\it $^1$ Stanford Institute for Theoretical Physics, Stanford University, Stanford, CA 94305}\\
{\it $^2$ SLAC National Accelerator Laboratory, 2575 Sand Hill, Menlo Park, CA 94025, USA.} \\

\end{center}

\vspace{1cm}

\begin{abstract}

We derive a simple relation between the Mellin amplitude for AdS/CFT correlation functions and the bulk S-Matrix in the flat spacetime limit, proving a conjecture of Penedones.   As a consequence of the Operator Product Expansion, the Mellin amplitude for any unitary CFT must be a meromorphic function with simple poles on the real axis.  This provides a powerful and suggestive handle on the locality vis-a-vis analyticity properties of the S-Matrix.  We begin to explore analyticity by showing how the familiar poles and branch cuts of scattering amplitudes arise from the holographic description.  For this purpose we compute examples of Mellin amplitudes corresponding to 1-loop and 2-loop Witten diagrams in AdS.  We also examine the flat spacetime limit of conformal blocks, implicitly relating the S-Matrix program to the Bootstrap program for CFTs.  We use this connection to show how the existence of small black holes in AdS leads to a universal prediction for the conformal block decomposition of the dual CFT.

\end{abstract}

\bigskip

\end{titlepage}
\section{Introduction and Review}

The only exact observables in quantum gravity are associated with the boundary of spacetime, so a holographic theory of asymptotically flat spacetime will be a theory of the S-Matrix.  Recently, evidence \cite{Penedones:2010ue, Fitzpatrick:2011ia, Paulos:2011ie} has accumulated supporting a conjecture of Penedones \cite{Penedones:2010ue} that  the Mellin representation \cite{Mack, MackSummary, Penedones:2010ue} of Conformal Field Theory (CFT) correlation functions defines a dual bulk S-Matrix via the vanishing-curvature limit of the AdS/CFT correspondence \cite{Maldacena, Witten, GKP}. 
Our goal is to investigate the Mellin amplitude as a way of organizing AdS/CFT into a holographic theory of flat spacetime where the  principles of locality and unitarity are evident.  The purpose of this work is to prove Penedones' conjecture \cite{Penedones:2010ue, Fitzpatrick:2011ia} and to explore how the locality vis-a-vis analyticity of the S-Matrix emerges from the Mellin amplitude of the CFT; we will discuss unitarity as a consequence of the OPE in a separate work \cite{Unitarity}.

The Mellin amplitude for a CFT correlator represents the correlator in terms of a set of variables that make CFT physics more manifest, much as momentum space naturally represents scattering amplitudes.    In the case of scalar operators, the Mellin representation  \cite{Mack, MackSummary, Penedones:2010ue} takes the form
\be
\left\langle \CO_1(x_1)... \CO_n(x_n) \right\rangle = \int [d \delta_{ij}] M(\delta_{ij}) \prod_{i<j}^n \Gamma(\delta_{ij}) (x_{ij})^{-2\delta_{ij}} 
\label{eq:mellin}
\ee
where the Mellin Amplitude is the conformally invariant function $M(\delta_{ij})$.  The `Mellin space' is the space of variables $\delta_{ij}$, which are symmetric and subject to the $n$ constraints $\sum_{j \neq i} \delta_{ij} = \Delta_i$, where $\Delta_i$ are the dimensions of the operators $\CO_i$.  One can always think of the $\delta_{ij}$ in analogy with the Mandelstam invariants $s_{ij}= 2 p_i \cdot p_j $ of a scattering amplitude; the constraints on the $\delta_{ij}$ are analogous to the constraints on the $s_{ij}$ that follow from momentum conservation and the on-shell conditions.

The most significant principle behind the simplicity of the Mellin amplitude is unitarity, in the guise of the Operator Product Expansion.  Using the convergent OPE, it is possible to express correlation functions involving many operators as a sum over products of correlators involving fewer operators.  The Mellin amplitude displays these OPE decompositions of correlation functions as a sum over poles, with residues given by lower-point Mellin amplitudes.  These poles are analogous to the multi-particle factorization channels of scattering amplitudes, but in fact they are even more constrained: in unitary CFTs the Mellin amplitude can only have simple poles in a certain region of the real axis.  The existence of these factorization channels makes it possible to determine the Mellin amplitude recursively, and we \cite{Fitzpatrick:2011ia} and Paulos \cite{Paulos:2011ie}  gave algebraic diagrammatic rules for its evaluation in the case of scalar theories at tree-level.  We will review these and other properties of Mellin amplitudes more thoroughly in section \ref{sec:Review}.

The similarity between the Mellin amplitude and the S-Matrix is no coincidence.  Penedones has conjectured \cite{Penedones:2010ue} a very simple and natural relationship between the Mellin amplitude of a large $N$ CFT and the S-Matrix of its bulk dual in the flat spacetime limit of AdS.  His conjecture takes the form
\be \label{eqn:PenedonesConjectureIntro}
T(s_{ij}) =  \lim_{R \to \infty} \frac{1}{\NN}  \delta^{d+1} \left( \sum_i p_i \right)\int_{-i\infty}^{i\infty} d \alpha \ \!  e^{\alpha}  \alpha^{h - \Delta_\Sigma}  
M \left( \delta_{ij} = -\frac{R^2 s_{ij}}{4 \alpha}, \Delta_a = R m_a \right)
\ee
where $R$ is the AdS length scale, $\NN$ is a normalization factor, and $T(s_{ij})$ is the connected, reduced S-Matrix of massless scalar particles as a function of the Mandelstam invariants.  Penedones checked this conjecture for general scalar theories at tree-level and for $\phi^4$ theory at one-loop.  Recently we \cite{Fitzpatrick:2011ia}  verified the conjecture for $n$-pt amplitudes in general scalar theories at tree-level by showing that our diagrammatic rules for the Mellin amplitude reduce to the usual Feynman rules in the flat space limit.  By setting up the appropriate scattering experiment \cite{Polchinski, Susskind, GGP, Fitzpatrick:2011jn} in AdS and making gratuitous use of the stationary phase approximation, we will derive Penedones' conjecture in section \ref{sec:FlatSpaceLimit}. 

\begin{figure}[t!]
\begin{center}
\includegraphics[width=0.95\textwidth]{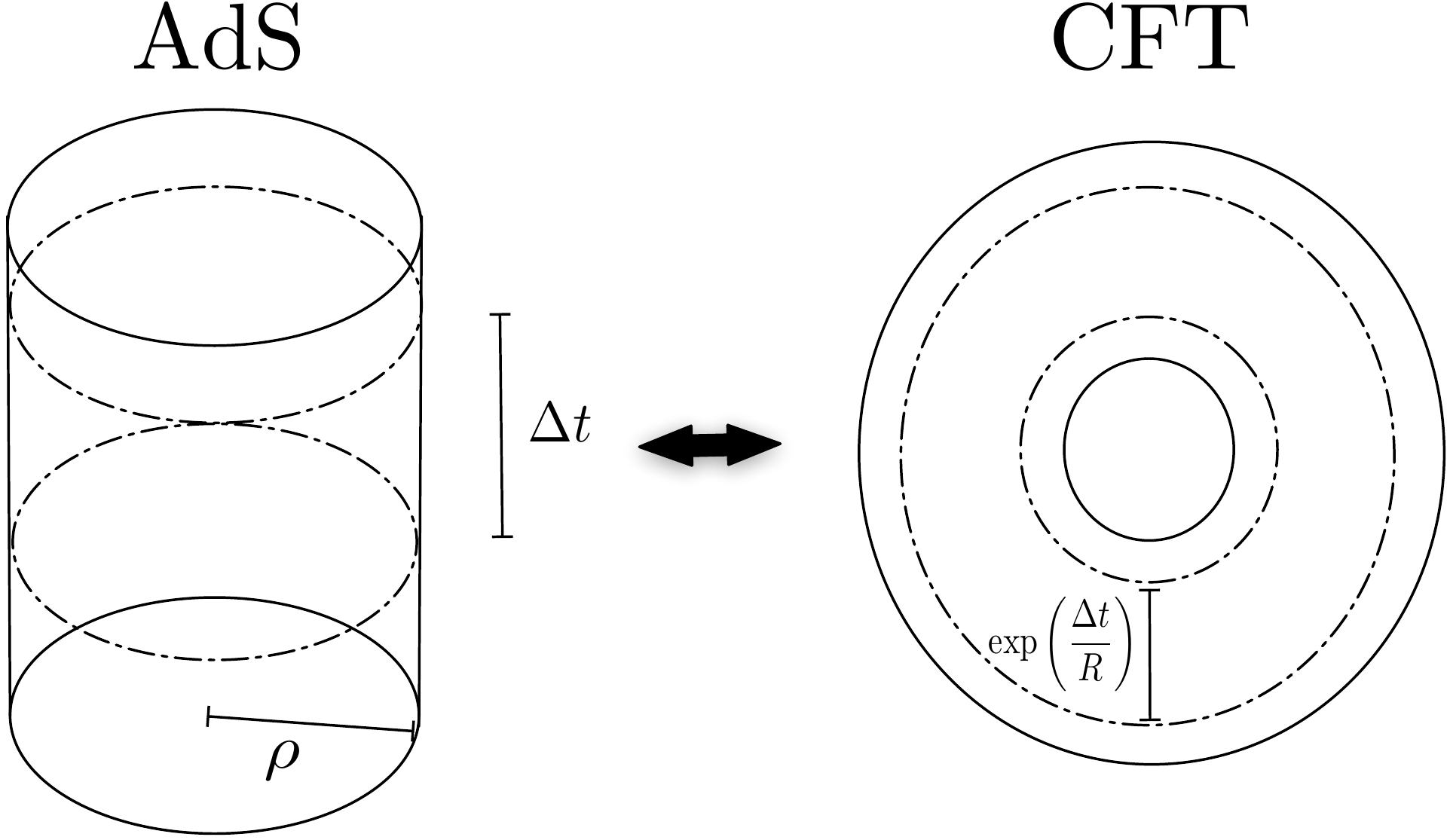}
\caption{ This figure shows how the AdS cylinder in global coordinates corresponds to the CFT in radial quantization.  The time translation operator in the bulk of AdS is the Dilatation operator in the CFT, so energies in AdS correspond to dimensions in the CFT.
\label{fig:AdSCylinderIntro}  }
\end{center}
\end{figure}

Why is the Mellin amplitude related to the flat-space S-matrix?  A key point is that the Dilatation operator in the CFT generates global time translations in AdS, as pictured in figure \ref{fig:AdSCylinderIntro}.  In other words, the energy of particles in AdS is given by the dimension of a CFT operator (or equivalently a CFT state through the operator-state correspondence).  So aside from their manifest similarity in a large class of examples, one can understand the relationship between Mellin and scattering amplitudes by thinking about which states in the CFT correspond to scattering processes in AdS.  
The CFT states dual to AdS particles with energies parametrically larger than the AdS curvature scale correspond to operators with very large scaling dimension.  The $\delta_{ij}$'s in the Mellin amplitude correspond to relative scaling dimensions, so scattering states localize the Mellin amplitude on large $\delta_{ij}$'s related to the energy and momentum of the physical scattering process.  We will show how to make this argument precise by directly extracting scattering states from the correlation functions of $n$ single-trace operators, written in the Mellin representation.  For scattering momenta $p_i$ large compared to the AdS curvature scale, the $\delta_{ij}$ integration variables in equation (\ref{eq:mellin})  are almost completely supported in the region where they line up with the Mandelstam invariants, so that we have $\delta_{ij} \propto p_i \cdot p_j$.  Thus, assuming only that single-particle states exist in the flat-space limit of the bulk theory, this gives a direct, non-perturbative proof of Penedones' conjecture, as stated in equation (\ref{eqn:PenedonesConjectureIntro}).

With Penedones' conjecture proven, we turn to an investigation of locality.  From the point of view of the S-Matrix, one must view locality\footnote{ See \cite{ArkaniHamed:2007ky, Giddings:2009gj} for nice and relevant discussions.} as the statement that the scattering amplitudes are exponentially bounded analytic functions of the kinematic invariants, except near the real axis, where the amplitudes can have poles and branch cuts corresponding to resonances and multi-particle states.  Analyticity of the bulk S-matrix is a particularly important property because, in contrast to most other criteria for locality, it is a sharp statement phrased directly in terms of holographic observables.  The fact that the Mellin amplitude must be a meromorphic function with simple poles on the real axis \cite{Mack, MackSummary} seems to us very striking: since the S-Matrix is just a simple transform of the Mellin amplitude, perhaps bulk locality can be understood as a natural consequence of a few simple conditions on CFTs.  

We will begin to explore these issues by demonstrating how various familiar physical analyticity properties such as branch cuts and resonances emerge from properties of the Mellin amplitude at loop-level in AdS.  
The emergence of simple poles is a straightforward application of equation (\ref{eqn:PenedonesConjectureIntro}) to the Mellin amplitude for a single tree-level exchange in AdS.  Indeed, the Mellin amplitude in this case already has an infinite series of poles, corresponding to the primary CFT operator
and its descendants, and all that is necessary is for them to coalesce into a single pole at the physical bulk mass.
However, how branch cuts and resonances emerge is less obvious, since the only singular behavior of the Mellin amplitude
is the presence of poles on the real axis.  The key element in this case is the presence of a tower of poles from multi-trace operators,
each of which contributes a pole to the holographic S-matrix.  In the flat-space limit, these poles appear to be arbitrarily
narrowly spaced, and we will see that from a line of such poles a branch cut emerges. 
Once branch cuts arise, resonances are virtually guaranteed: the resummation of one-particle irreducible (1PI) Witten diagrams
brings the loop contribution into the denominator of the propagator. We will show in more detail that the
mechanism for turning real poles in the Mellin amplitude into poles with large imaginary components
is inherently connected to the coalescence of poles from multi-trace operators contributing to 1PI diagrams.  All of these issues are studied in section \ref{sec:Examples}, where we also comment on the significant recent progress \cite{JP, Katz, Heemskerk:2010hk, Heemskerk:2010ty} that we have seen in understanding locality in AdS/CFT.

In section \ref{sec:ConformalBlocksandBH} we study the relationship between the conformal block decomposition of the CFT and the flat space S-Matrix.  Conformal blocks are the most elementary possible contributions to CFT 4-pt functions consistent with conformal symmetry; they correspond to the exchange of a single primary operator in the CFT and are nicely reviewed in \cite{Rattazzi:2008pe}.
Since a conformal block has definite dimension and angular momentum, in the flat space limit we find that it is simply a delta function in the center-of-mass energy times the appropriate angular-dependence for a spin $\ell$ state:
\ba
B_\Delta^\ell (\delta_{ij}) &\leadsto& P_\ell^{(d)} (\cos \theta) \delta(s-M^2)  .
\ea
Recently, progress \cite{Dolan:2003hv, Rattazzi:2008pe, Costa:2011mg, Costa:2011dw, Hellerman:2009bu, Poland:2011ey, Rychkov:2011et}  has been made understanding general properties of CFTs using only the conformal block decomposition, unitarity, and crossing symmetry.  Via the flat spacetime limit, this means that there is a natural relationship between this version of the Bootstrap program \cite{Belavin:1984vu} and the S-Matrix program.

Finally, we  discuss what the presence of flat-space black holes in bulk scattering amplitudes can tell us about CFT correlation functions.  The behavior of high-energy scattering amplitudes directly tells us about the size of contributions from
conformal blocks of large dimension.  In theories with gravity, high-energy scattering at trans-Planckian
energies results in black hole formation, which suggests very rapid shutdown of the scattering amplitude
for 2-to-2 scattering
\be
\mathcal{S}(s) \sim \exp \left[-\frac{1}{2} S_{BH}(s) \right] 
\ee
due to the large entropy of the black hole.  This shutdown of the amplitude corresponds to a rapid shutdown of the conformal block decomposition at large dimensions: we expect that the conformal block decomposition of CFTs will truncate at some large dimension  to more than  exponentially good accuracy, as we discuss in section \ref{sec:BH}.  Deriving this property directly from the CFT would involve understanding the microphysics of Hawking evaporation, and thus seems to be an extremely important goal for future work.

Since our primary object of interest will be the flat space S-Matrix, the vast majority of our discussions of CFTs will be in the context of large $N$ CFTs with an AdS dual that is well-described by effective field theory.  We will always discuss the correspondence in terms of AdS$_{d+1}$/CFT$_d$, and ignore any compactification manifolds, although we expect that they will be straightforward to include in the future.  
In the following subsections, we will discuss and review some basics of CFTs and the Mellin amplitude, and provide a new, simplified derivation of the vertices for the diagrammatic rules discovered in \cite{Fitzpatrick:2011ia, Paulos:2011ie}.

\subsection{Review}
\label{sec:Review}

\subsubsection{Space of States at Large $N$}

The Mellin representation of CFT correlation functions is best understood by thinking of CFT correlators in analogy with flat space scattering amplitudes.  The basis of this analogy is the Hilbert space of states at large $N$.  In flat space, one typically quantizes the theory on flat,  space-like surfaces, and chooses the  Hamiltonian to be the generator of Poincar\'e time translations. One then takes the spectrum of this Hamiltonian to be the basis of the space of states, which in the limit of weak interactions are the single- and multi-particle states and form a Fock space.  This space has the essential physical property that the energy of a multi-particle state is simply the sum of the energies of the individual particles.  This description remains valid even in cases where the  underlying Lagrangian of the theory is strongly coupled, as long as the physical particles of the theory are weakly interacting.

CFTs at large $N$ have a nearly identical structure when one studies the theory in radial quantization, rather than the quantization in Poincar\'e time.  That is, one takes the ``Hamiltonian'' of the theory to be the dilatation operator $D = i x^a \frac{\partial }{\partial x^a}$, which generates  radial evolution by rescaling.  Rather than having a definite frequency, eigenstates of the ``Hamiltonian'' now have a definite scaling dimension.
Such eigenstates no longer correspond to particles per se, which anyway do not exist due to the scale-invariance of the theory, but rather to operators, through the state-operator correspondence of radial quantization.  Simply put, the state $|\CO\>$ corresponding to an operator $\CO$ is just that operator acting at the origin on the vacuum: $| \CO \> = \CO(0) | \textrm{vac} \>$.  The assumption of large $N$ can be stated purely in terms of the gauge-invariant operators of the theory as a dynamical property of their correlation functions: at $N\rightarrow \infty$, there is a subsector of operators of the theory whose two-point functions completely determine all correlation functions.  For instance, the four-point function of such operators is just the sum of disconnected pieces:
\be
\< \CO_1(x_1) \CO_2(x_2) \CO_3(x_3) \CO_4(x_4) \> \stackrel{N\rightarrow \infty}{=}
 \< \CO_1(x_1) \CO_2(x_2) \> \< \CO_3(x_3) \CO_4(x_4) \> + (2\leftrightarrow 3) + (2 \leftrightarrow 4) .
 \ee
 These operators are called the ``single-trace'' operators, and they create CFT states that are the analogs of single-particle states in Poincar\'e QFT, because in the large $N$ limit, the multi-trace states are just a direct product of single-trace states. It is useful and conventional to group all operators into irreducible representations of the conformal group, the lowest-dimension states of which are the ``primary'' operators.  Primary operators are physically similar to flat-space particle states with zero total momentum, and knowledge of their correlation functions is sufficient to determine all correlation functions of the theory.
 The spatial dependence of their two-point functions is completely constrained by conformal symmetry, and we will work with the following normalization
 for single-trace scalar primaries:
 \ba
 \< \CO(x) \CO(0) \> &=& \frac{ \CC_\Delta}{x^{2\Delta}}, \ \ \ \ \ \CC_\Delta = \frac{\Gamma(\Delta)}{2 \pi^h \Gamma(\Delta+1-h)}
 \ea
where $h=\frac{d}{2}$.     Dimensions of multi-trace states are given by the sum of the individual single-traces.  In other words, given two operators $\CO_1$ and $\CO_2$ with dimensions $\Delta_1$ and $\Delta_2$, then at infinite $N$ there exists an operator ``$\CO_1 \CO_2$'' with dimension $\Delta_1 + \Delta_2$, and furthermore all operators can be written as a product of the single-traces.  The full space of states is again just a Fock space, the single-trace operators giving rise to the individual creation and annihilation operators of the theory.  We will next review
 how it is that Mellin space makes this structure manifest in the correlation functions.

\subsubsection{Factorization on Poles in the Mellin Amplitude}

Specific operators appear in the Mellin amplitude in a particularly transparent way: as poles in the Mellin variables $\delta_{ij}$.  The reason is that operators are ``exchanged'' in correlation functions whenever they appear in the operator product expansion (OPE) of the external operators, e.g. for scalars
\be
 \CO_1(x_1) \CO_2(x_2) \supset c_{12 \CO} x_{12}^{\Delta_\CO -\Delta_1 - \Delta_2}  \CO(x_2) , \ \ \ \ \CO_3(x_3) \CO_4 (x_4) \supset c_{34 \CO}x_{34}^{\Delta_\CO - \Delta_3 - \Delta_4}  \CO (x_4), 
 \ee
 and this indicates that correlation functions contain a contribution with a particular scaling:
\be
  \< \CO_1(x_1) \CO_2(x_2) \CO_3(x_3) \CO_4(x_4) \>\supset c_{12 \CO} c_{34 \CO} \left(\frac{x_{24}}{x_{14}} \right)^{\Delta_1 - \Delta_2} \left( \frac{x_{14}}{x_{13}} \right)^{\Delta_3 - \Delta_4} \frac{u^{\frac{\Delta_\CO}{2} } f(v)
 }{x_{12}^{\Delta_1 + \Delta_2} x_{34}^{\Delta_3 + \Delta_4} } , 
 \ee
 \be
  u = \left( \frac{x_{12}^2 x_{34}^2 }{ x_{24}^2 x_{13}^2 } \right), \ \ \ v = \left( \frac{x_{14}^2 x_{23}^2 }{ x_{24}^2 x_{13}^2 } \right),
\ee
where we have used the fact that the scalar four-point function must take the form $ \frac{F(u,v)}{x_{12}^{\Delta_1+\Delta_2} x_{34}^{\Delta_3+\Delta_4}}$ $\times$ $\left(\frac{x_{24}}{x_{14}} \right)^{\Delta_1 - \Delta_2} \left( \frac{x_{14}}{x_{13}} \right)^{\Delta_3 - \Delta_4}$ by conformal invariance.  Meanwhile, the contour integration over $\delta_{ij}$ variables in the Mellin integral causes them to become localized on the poles, which in this case reproduces the correct power of $u$ when there is a pole in the Mellin amplitude at
\be
\delta_{LR} \equiv \Delta_1 + \Delta_2 - 2 \delta_{12} = \Delta_\CO.
\ee
It is often convenient to introduce associated with each external operator $\CO_i(x_i)$  a fictitious ``momentum'' $p_i$ subject to the ``on-shell'' and 
``momentum-conservation'' condtions $p_i^2 = \Delta_i$ and $\sum_i p_i =0$.  Then, the constraints $\sum_{j \ne i} \delta_{ij} = \Delta_i $
are automatically satisfied formally by the inner products $\delta_{ij} \equiv - p_i \cdot p_j$.  Furthermore, the poles associated with a 
given OPE are in exactly the linear combination of $\delta_{ij}$'s corresponding to the fictitious momentum-squared flowing through
the corresponding channel.  For instance, note that the $\delta_{LR}$ variable above is simply $\delta_{LR} =-(p_1 + p_2)^2 = -(p_3 + p_4)^2$.  
This fact, combined with the Fock space nature of the states at infinite $N$, is the reason for the $\Gamma$ functions in the Mellin integrand
in eq. (\ref{eq:mellin}).  The point is that these $\Gamma$ functions contain poles when the $\delta_{ij}$ are negative integers,
which encodes the presence of multi-trace operators in the theory.  For instance, the pole in $\Gamma(\delta_{12} )$ at $\delta_{12}= -m$ 
is a pole in $\delta_{LR}$ at $\delta_{LR} = \Delta_1 + \Delta_2 + 2 m$, which corresponds to double-trace operators of the form
$\partial^r  \CO_1 \partial^{m-r} \CO_2$. In a perturbative description in $1/N$, it is natural to factor out these poles in the integrand from what one calls the ``Mellin amplitude'' $M(\delta_{ij})$, as this separates out universal purely ``kinematic'' information.  

One of the most important features of the Mellin amplitude is that it factors on these poles into the Mellin amplitudes corresponding to lower-point correlation functions.  In \cite{Fitzpatrick:2011ia} it was shown that this factorization takes the following form for the exchange
of a scalar field in AdS dual to an operator of dimension $\Delta_\CO$:
\ba
Res [ M_n(\delta_{ij}) ]_{\delta_{LR} = \Delta_\CO +2m}&=& -4 \pi^h \sum_m  \frac{\Gamma(\Delta-h+1) m!}{(\Delta-h+1)_m} L_m(\delta_{ij}) R_m(\delta_{ij}), \nn\\
L_m(\delta_{ij}) &=& \sum_{\sum n_{ij} = m }  M^L_{k+1} (\delta_{ij} + n_{ij}) \prod_{i<j}^k \frac{(\delta_{ij})_{n_{ij}} }{n_{ij}!} , \nn\\
R_m(\delta_{ij}) &=& \sum_{\sum n_{ij} = m }  M^R_{n-k+1} (\delta_{ij} + n_{ij}) \prod_{k<i<j}^n \frac{(\delta_{ij})_{n_{ij}} }{n_{ij}!} , 
\ea
where $k $ and $n-k$ are the number of operators to the left and right, respectively of the scalar propagator, and $M^L_{k+1}, M^R_{n-k+1}$ are
the Mellin amplitudes for the corresponding $(k+1)$- and $(n-k+1)$-point correlation functions obtained by cutting that propagator.

\subsubsection{Finite Difference Equation}

Many properties of the Mellin amplitudes for Witten diagrams in AdS can be understood most easily by using a finite difference equation that encodes bulk propagation and conformal blocks.  Its existence follows from the fact that propagators are solutions of the bulk Klein-Gordon equation with a $\delta$-function source, e.g.
\be
[ \nabla^2_{\rm AdS} - \Delta (\Delta-d)] G_\Delta(X,Y) = - \delta(X,Y),
\ee
for scalars, where $\delta(X,Y)$ is the $\delta$-function corresponding to the covariant measure in AdS.  Furthermore,
one can act with the conformal Casimir on all the external operators inserted to the left side of the propagator in the Witten 
diagram, and this is equivalent to the Laplacian acting on the bulk propagator : $\frac{1}{2} (\sum_{i \in L} J_i)^2 \cong
  - \nabla_{\rm AdS}^2$. One then obtains a differential equation in position space for the correlation function,
  \ba
  \left[ \frac{1}{2} \left( \sum_{i \in L} J_i \right)^2 - \Delta( d-\Delta) \right] A &=& A_0,
  \ea
  where $A$ is the correlation function itself, and $A_0$ is the correlation function for the same Witten diagram but with the
  propagator replaced by a contact interaction.   One major advantage of Mellin space is that the differential
  equation in position space becomes a finite difference equation in Mellin space, in much the same way that the Klein-Gordon equation becomes algebraic in momentum space.  For any number of external legs, this takes the form
    \ba \label{eqn:FunctionalEquation}
  M_0 &=& (\delta_{LR} - \Delta) (d - \Delta - \delta_{LR}) M + \sum_{ab \le k < ij} 
  \left( \delta_{ai} \delta_{bj} M - \delta_{aj} \delta_{bi} M_{ai,bj}^{aj,bi} + \delta_{ab} \delta_{ij} M^{ab,ij}_{ai,bj} \right),
  \ea
  where $\delta_{LR} = - \left( \sum_{i \in L} p_i \right)^2 = \sum_{i \in L } \Delta_i - 2 \sum_{ i<j \in L} \delta_{ij}$, 
  $M^{ab, ij}_{ai,bj} = M(\delta_{ab}+1, \delta_{ij} +1, \delta_{ai}-1,\delta_{bj}-1)$, and $M_0$ is the Mellin amplitude
  for $A_0$.  The reason for the finite differences is that conformal invariance determines the contribution from any descendant of a primary operator $\CO_p$ in terms of the contribution of the primary itself, and the dimensions  of the descendants differ from those of their primaries by integers.  Since the conformal blocks are just eigenfunctions of the conformal Casimir, they satisfy this same equation with $A_0 = M_0 =  0$, as we discuss in section \ref{sec:ConformalBlocksandBH} and derive in appendix \ref{sec:ConformalBlocks}. 

\subsubsection{A Simplified Derivation of Vertices in the Diagrammatic Rules}
\label{sec:DiagrammaticRules}

\begin{figure}[t!]
\begin{center}
\includegraphics[width=0.40\textwidth]{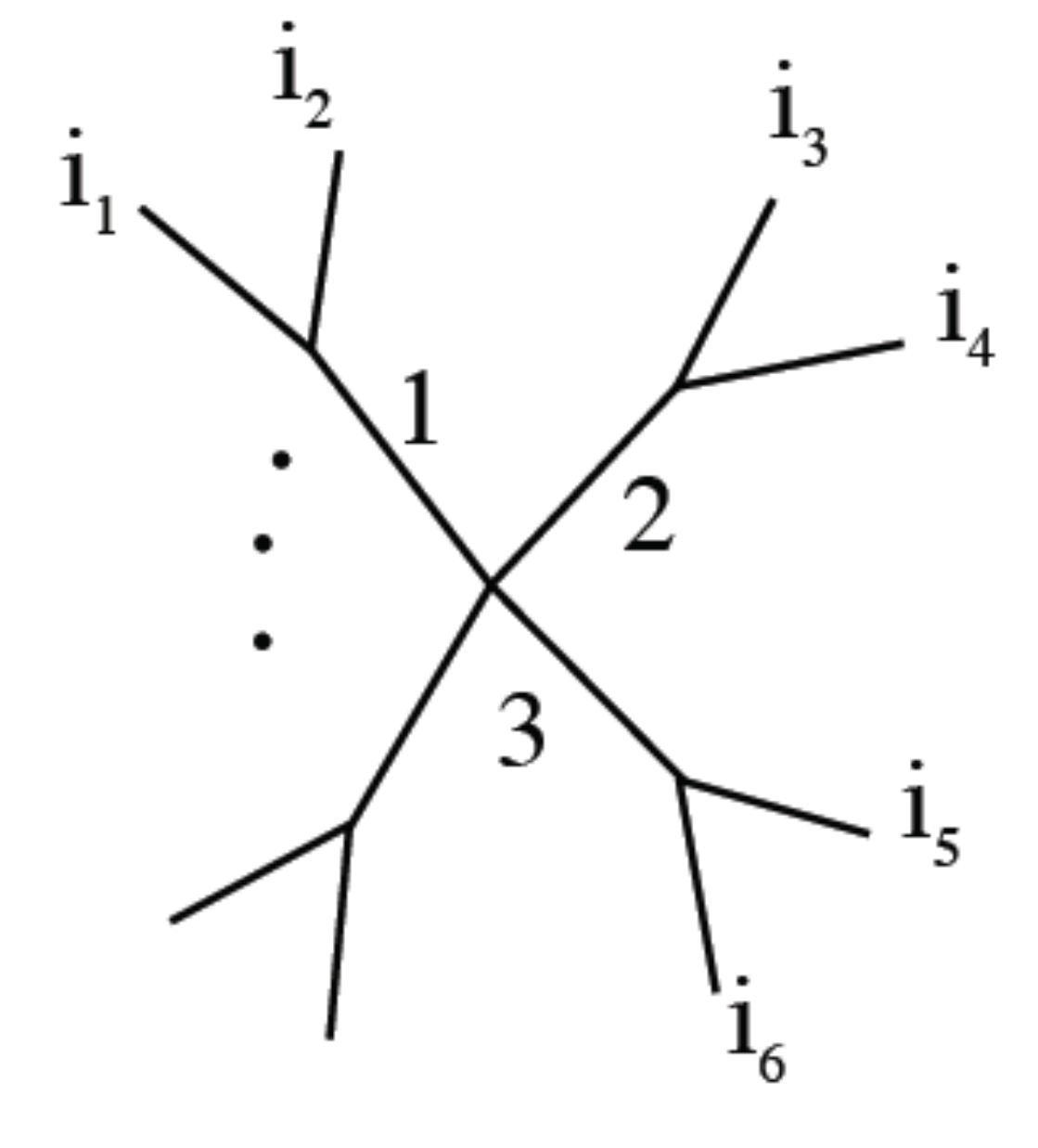}
\caption{Internal vertex described in the text.  Internal vertices with off-shell $\delta_{ij}$'s flowing through them
can be obtained by adding on external three-point vertices.
\label{fig:simplevertex}  }
\end{center}
\end{figure}

It was recently discovered \cite{Fitzpatrick:2011ia, Paulos:2011ie} that there are simple diagrammatic rules that rapidly determine the amplitude for any tree-level Witten diagram in Mellin space. In \cite{Paulos:2011ie} very simple and general formulas for these rules were discovered empirically  and conjectured to hold more generally, whereas in \cite{Fitzpatrick:2011ia} equivalent but more cumbersome results were proven. Here, we will streamline the derivation of the form of the vertex factors in the diagrammatic rules through a use of the finite difference equation that we reviewed in the previous subsection.   Let us consider the diagram that connects $n$ 3-pt vertices to an off-shell $n$-pt vertex in the middle, as depicted in figure \ref{fig:simplevertex}.  We will assume that the diagram therefore only depends on the variables $2 \delta_i = \Delta_{i_a} + \Delta_{i_b} - 2 \delta_{2i-1 \ \! 2i} - \Delta_i = 2 \Delta_{i_a i_b,i} - 2 \delta_{2i-1 \ \!  2i}$ and see that this is a consistent assumption. As reviewed in the previous subsection, by acting with the conformal Casimir on the external operators at $i_1$ and $i_2$, one can collapse the internal bulk propagator $\delta_1$ to obtain a finite difference equation, which in this case reads 
\be
0 = (\delta_{LR} - \Delta_1)(d - \Delta_1 - \delta_{LR}) M + 2 \delta_{12} \sum_{ij > 2}  \delta_{ij} M^{12,ij} .
\ee
Since $M$ depends only on the $\delta_i$ combinations, we can group together $\delta_{ij}$'s into the $\delta_i$ linear combinations
in the above equation:
\bea
0 &=& 2\delta_1(d - 2\Delta_1 - 2 \delta_1) M + 4(\Delta_{i_a i_b ,i} - \delta_1) \sum_{i \geq 2}  (\Delta_{i_a i_b,i} - \delta_{i}) M^{12; 2i-1 \ \! 2i} \nonumber \\
&& + 2(\Delta_{1_a 1_b,1} - \delta_1) \left(\Delta_1 + \sum_{i=1}^n 2\delta_i - \sum_{j=2}^n 4\Delta_{j_a j_b,j}  \right) M^{12} .
\eea
Now, starting with the assumption that there exists a set of diagrammatic rules for the Mellin amplitude in figure \ref{fig:simplevertex}, we will derive the explicit form that the vertex factors of those rules have to take.  So we take the following ansatz for the Mellin amplitude:
\be
M = \sum_{m_i} \frac{X_{m_1... m_n}}{\delta_i - m_i} = \sum_{m_i} V_{m_1... m_n} \prod_j^n \frac{S(m_j) V_3^j(m_j)}{\delta_i - m_i},
\ee
where $V_{m_1...m_n}$ is an off-shell n-pt vertex, and we have suppressed dependence on the various dimensions.  The function 
\be
S(m_i) = \frac{-2 \pi^h \Gamma^2(\Delta_i -h+1) m_i!}{ \Gamma(\Delta_i + m_i  - h +1)}
\ee
has been associated with Mellin space propagators by convention \cite{Fitzpatrick:2011ia, Paulos:2011ie}, although it could also be absorbed into the definition of the vertices.  The reduced vertices $V_3^j(m_j)$ are vertex factors associated with two external lines and a single internal line, and take the form
\ba
V_3^j(m_j) &=& \frac{\lambda_{i_a i_b j} }{m_j! (\Delta_{i_a i_b, j})_{-m_j}} ,
\ea
where the coefficient $\lambda_{i_a i_b j}$ is just the Mellin amplitude for the corresponding three-point function.\footnote{
For  example,  this is $\lambda_{i_a i_b j}= g\frac{\pi^h}{2} \Gamma\left(\frac{\Delta_{i_a} + \Delta_{i_b} +\Delta_j}{2} -h\right) \prod_{i=i_a,i_b,j} \CC_{\Delta_i}$ for a $g \phi_{i_a} \phi_{i_b} \phi_j$ interaction, in our normalization.}  
The object $X$ is just the total numerator, defined to simplify notation.  Plugging this in and evaluating on the poles we find
\bea
0 &=& \sum_{m_i} \left[ -4m_1 (\Delta_1 + m_1-h) X_{m_1... m_n}  + 4 (\Delta_{1_a,1_b,1} - m_1) \sum_{i \geq 2}  (\Delta_{i_a i_b ,i} - m_{i}) X_{m_1 - 1,...m_{i-1},m_i-1,m_{i+1}...,m_n}  \right. \nonumber \\
&& + \left. 2 ( \Delta_{1_a 1_b,1} -  m_1) \left(-\Delta_1 - 2m_1 + \sum_{i=2}^n [ 2m_i + \Delta_i ]   \right)  X_{m_1-1,...} \right].
\eea
 Now if we divide by the 3-pt vertices and propagators, the dependence on the external dimensions goes away!  This follows because
 \be
 \frac{S(m_j-1) V_3^j(m_j-1)}{S(m_j) V_3^j(m_j)} = \frac{\Delta_j +m_j-h}{\Delta_{j_a j_b, j} -m_j } .
 \ee
 Thus after dividing through, we find a simpler recursion relation for $V$
\bea
m_1 V_{m_1... m_n}  = \sum_{i \geq 2} [ (\Delta_i - h + m_i)  V_{m_1 - 1,...,m_i-1,...,m_n} ] +  \left(m_1 + \frac{\Delta_1}{2}  -\sum_{i=2}^n \left[ m_i + \frac{\Delta_i}{2} \right]   \right)  V_{m_1-1,m_2,...,m_n}  , \nonumber
\eea
where we assume that $m_1 > 0$.   Of course we also get similar equations from each of the $n$ legs, for a total of $n$ recursion relations. 
This formula thus provides a constructive derivation of $V$, since starting from $V_{0, \dots, 0}$, one may
recursively increase any of the $m_i$'s using the formula above; any $V$ with a negative index is to be understood to vanish.  One can
check that this recursion relation is satisfied by the compact expression for the vertex factors conjectured by Paulos \cite{Paulos:2011ie}, which are
\ba 
\label{eqn:Vertices}
 V_{m_1, \dots, m_n} &=& \lambda_n \left( \prod_{i=1}^n\frac{  (1-h+\Delta_i)_{m_i} }{m_i!}\right) \\
&& \times F_A^{(n)} \left( \Delta_\Sigma - h, \left\{     -m_1, \dots, -m_n \atop  1+\Delta_1 -h, \dots, 1+\Delta_n-h \right\} ; 1, \dots, 1 \right). \nn
 \ea
 The normalization of the vertex factors is chosen here to coincide with \cite{Fitzpatrick:2011ia}, and differs slightly from \cite{Paulos:2011ie}.  In the above expression, $\Delta_\Sigma \equiv \frac{1}{2} \sum_{i=1}^n \Delta_i$ , and $F_A^{(n)}$ is a Lauricella function:
 \ba
 F_A^{(n)} \left(g, \left\{ a_1, \dots, a_n \atop b_1, \dots, b_n \right\} ; x_1, \dots, x_n \right) &=& \sum_{n_i =0}^\infty \left(
 (g)_{\sum n_i} \prod_{i=1}^n \frac{(a_i)_{n_i}}{(b_i)_{n_i}} \frac{x_i^{n_i}}{n_i!} \right) .
 \ea
 The coefficient $\lambda_n$ is independent of $m_i$, and should be chosen so that $V_{0 \dots 0}$ is the on-shell $n$-point
 Mellin amplitude corresponding to the $n$-point vertex in figure \ref{fig:simplevertex}.  For $g \phi^n$,  
 \ba
 \lambda_n &=& g \frac{\pi^h}{2} \Gamma( \Delta_\Sigma- h) \prod_{i=1}^n \CC_{\Delta_i} .
 \ea

\emph{In summary}:  to compute any tree-level Witten diagram in a theory of scalars interacting via contact interactions, one draws all appropriate diagrams, just as in flat space-time.  To each vertex one associates a factor of the coupling times $V_{m_1,...,m_n}$, setting $m_i$ to zero for the external lines.  To each propagator one associates a factor of 
\be
\frac{S_{\Delta_i}(m_i) }{\delta_i - m_i}
\ee
where $\delta_i$ is the linear combination of $\delta_{ij}$ appropriate to a given propagator.  The combination $\delta_i$ can be most easily computed by associating fictitious momenta $p_i$ to each operator, computing $2 \delta_I = \left( \Sigma_i p_i \right)^2 - \Delta_I$, and setting $p_i^2 = \Delta_i$ and $p_i \cdot p_j = -\delta_{ij}$.  Finally, one sums over all $m_i$ associated with internal lines.

\begin{figure}[t!]
\begin{center}
\includegraphics[width=0.95\textwidth]{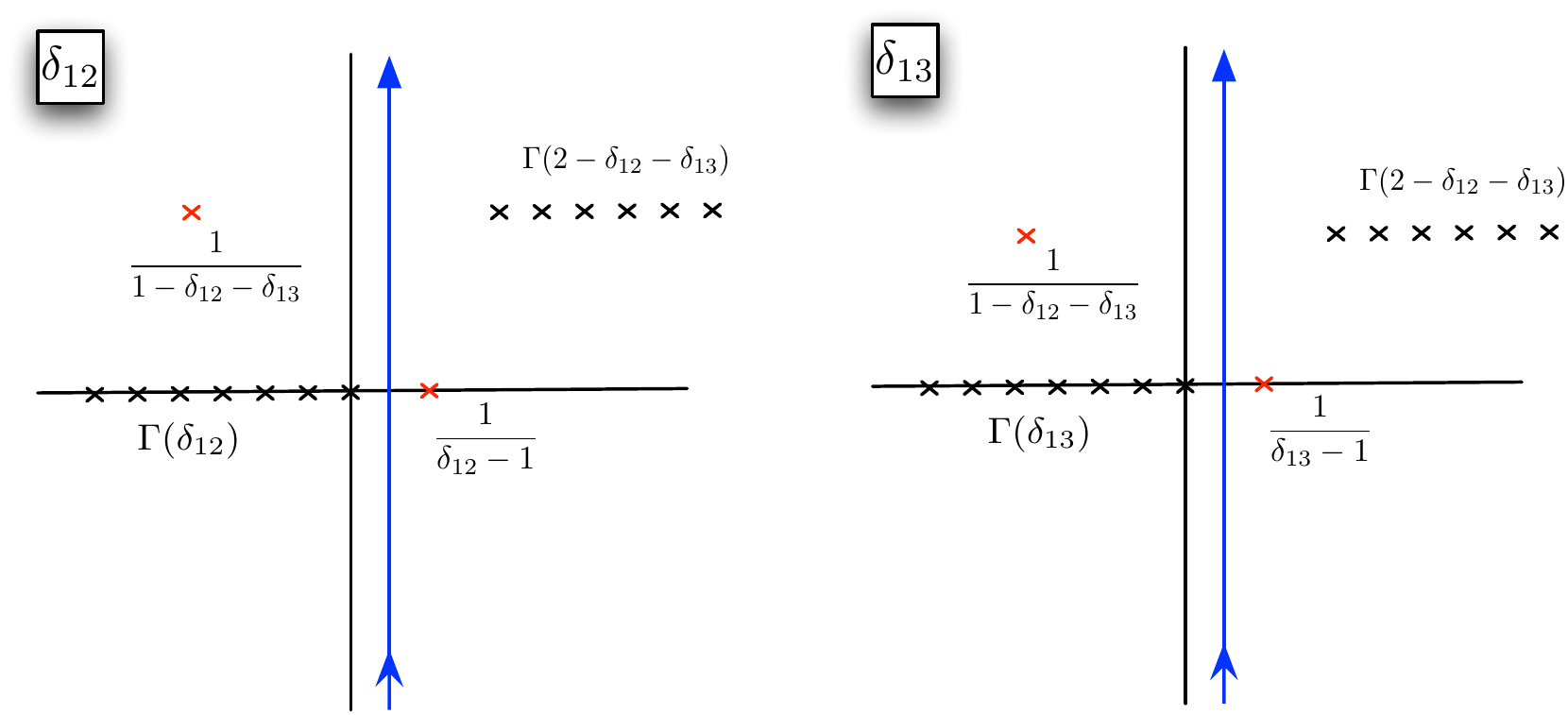}
\caption{ This figure shows the pole prescription for the contour integrals defining the Mellin representation, as a function of $\delta_{12}$ and $\delta_{13}$, for the very simple 4-pt example in equation (\ref{eqn:SimplestExample}).  One can see that the contour of integration lies between the poles of the $\Gamma$ functions (black) and the poles of the Mellin amplitude (red) in each variable, including $\delta_{14} = \Delta_\phi - \delta_{12} - \delta_{13}$.
\label{fig:PolePrescription}  }
\end{center}
\end{figure}

\subsubsection{A Simple Example - $\mu \phi^3$ Theory at Tree Level}

Let us now give the simplest possible example of a non-trivial Mellin amplitude.  If we take $d=\Delta_\phi=2$ and compute the Mellin amplitude corresponding to $\frac{1}{6}\mu \phi^3$ theory in AdS$_3$, we find
\be \label{eqn:SimplestExample}
M(\delta_{ij}) = \frac{R^3 \mu^2}{2 (4 \pi)^3}  \left(\frac{1}{\delta_{12}-1} + \frac{1}{\delta_{13}-1} + \frac{1}{\delta_{14}-1} \right)
\ee
The result is extremely simple because the (often infinite) sums in the definitions of the diagrammatic rules terminate for certain special values of the dimensions, as can be seen by inspection of equation (\ref{eqn:Vertices}).  Besides displaying for the reader how simple and natural Mellin amplitudes can be, this example affords an opportunity to see in figure \ref{fig:PolePrescription} the precise pole prescription for the contour integrals defining the CFT correlator in the Mellin representation.

\section{The Flat Space Limit}
\label{sec:FlatSpaceLimit}

Penedones has conjectured \cite{Penedones:2010ue} that the connected part of the S-Matrix of the bulk theory dual to a large $N$ CFT can be obtained from a simple integral transform of the Mellin amplitude
\ba
T(s_{ij}) &=&  \lim_{R \to \infty} \frac{1}{\NN}  \delta^{d+1} \left( \sum_i p_i \right) \int_{-i\infty}^{i\infty} d \alpha \ \!  e^{\alpha}  \alpha^{h - \Delta_\Sigma}  
M \left( \delta_{ij} = -\frac{R^2 s_{ij}}{4 \alpha}, \Delta_a = R m_a \right)
\label{FlatSpaceLimitFormula} \nn\\
\NN &=& \frac{\pi^h R^{\frac{n(1-d)}{2} + d +1} }{2} \prod_{i=1}^n \frac{\CC_{\Delta_i} }{\Gamma(\Delta_i )},
\ea
where we have introduced the short-hand symbol $\Delta_\Sigma = \frac{1}{2} \sum_i \Delta_i$ for half the sum of the external dimensions, and $\Delta_a$ are the dimensions of internal fields to which we wish to assign a non-zero mass in the flat space limit.   The integration contour in the $\alpha$ plane runs to the right of all poles of the Mellin amplitude, and the branch cut from $\alpha^{h-\Delta_\Sigma}$.   Penedones provided many pieces of evidence for equation (\ref{FlatSpaceLimitFormula}), showing that it works for tree-level and one-loop 4-pt amplitudes, and that it accords with earlier observations \cite{GGP} about a certain singularity in CFT correlators connected with flat space scattering amplitudes.  This evidence was further bolstered when it was shown in \cite{Fitzpatrick:2011ia, Paulos:2011ie} that Mellin amplitudes can be constructed directly from diagrammatic rules that reduce to the standard Feynman rules in the flat space limit.  In effect, this proved that equation (\ref{FlatSpaceLimitFormula}) is correct for all tree amplitudes in scalar field theories.

In what follows we will prove equation (\ref{FlatSpaceLimitFormula}) exactly for massless external scalar particles (internal `virtual' particles can be massive or massless) using the constructions of \cite{Polchinski, Susskind, GGP, Fitzpatrick:2011jn}.  The flat space S-Matrix can be extracted from AdS/CFT correlation functions in a straightforward manner; here we will only give a brief summary and refer the readers to \cite{Fitzpatrick:2011jn} for a thorough discussion.  A similar direct quantization of AdS fields was given by \cite{Banks:1998dd, Balasubramanian:1998de, Bena:1999jv} in the early days of AdS/CFT, and has been revisited in detail recently by \cite{Harlow:2011ke}.    

Individual particles are created by single-trace CFT Operators in the large $N$ limit.  We would like to prepare and then measure scattering states that correspond to many particles with definite energy and momentum in the center of AdS.  To create a massless particle in a plane wave state with energy $\omega$ and velocity $\hat v$ that passes through the center of AdS at time $t=0$, we act with the single-trace operator $\CO(t, \hat x)$ on the vacuum as
\be \label{eq:SingleParticleNorm}
|\omega, \hat v \rangle = \frac{2^{\Delta}\Gamma(\Delta)R^{\frac{d-3}{2}}}{(2\pi)^{h+1}\CC_\Delta (R \omega)^{\Delta-1}} \int_{-\frac{\pi R}{2} - \tau}^{-\frac{\pi R}{2} + \tau} dt
e^{i \omega t} \CO(t, -\hat v) |0 \rangle
\ee
where $\tau \ll R$ so that the in and out operators have non-overlapping support \cite{Gary:2009mi},  and $\Delta$ is the dimension of $\CO$.  The prefactor in front of the integral normalizes these states so that
 \be
 \langle \omega, \hat v | \omega', \hat v' \rangle = 2\omega \delta^d_\tau(\vec p + {\vec p} \ \! ') 
 \ee
 where the delta function is regulated by the length scale $\tau$ (we have included the derivation in appendix \ref{sec:SingleParticleNorms}).  Next we send $R \to \infty$ followed by $\tau \to \infty$ to take the flat space limit, keeping the physical energy $\omega$ fixed.  To prepare a multi-particle in-state one simply acts on the vacuum with several different single-trace operators.  One measures the out-states in an identical way, except replacing $- \frac{\pi R}{2} \to \frac{\pi R}{2}$, $\hat v \to - \hat v$, and taking the Hermitian conjugate.  The overlap between the in-states and the out-states extracts the S-Matrix for plane waves from a CFT correlation function.

\begin{figure}[t!]
\begin{center}
\includegraphics[width=0.95\textwidth]{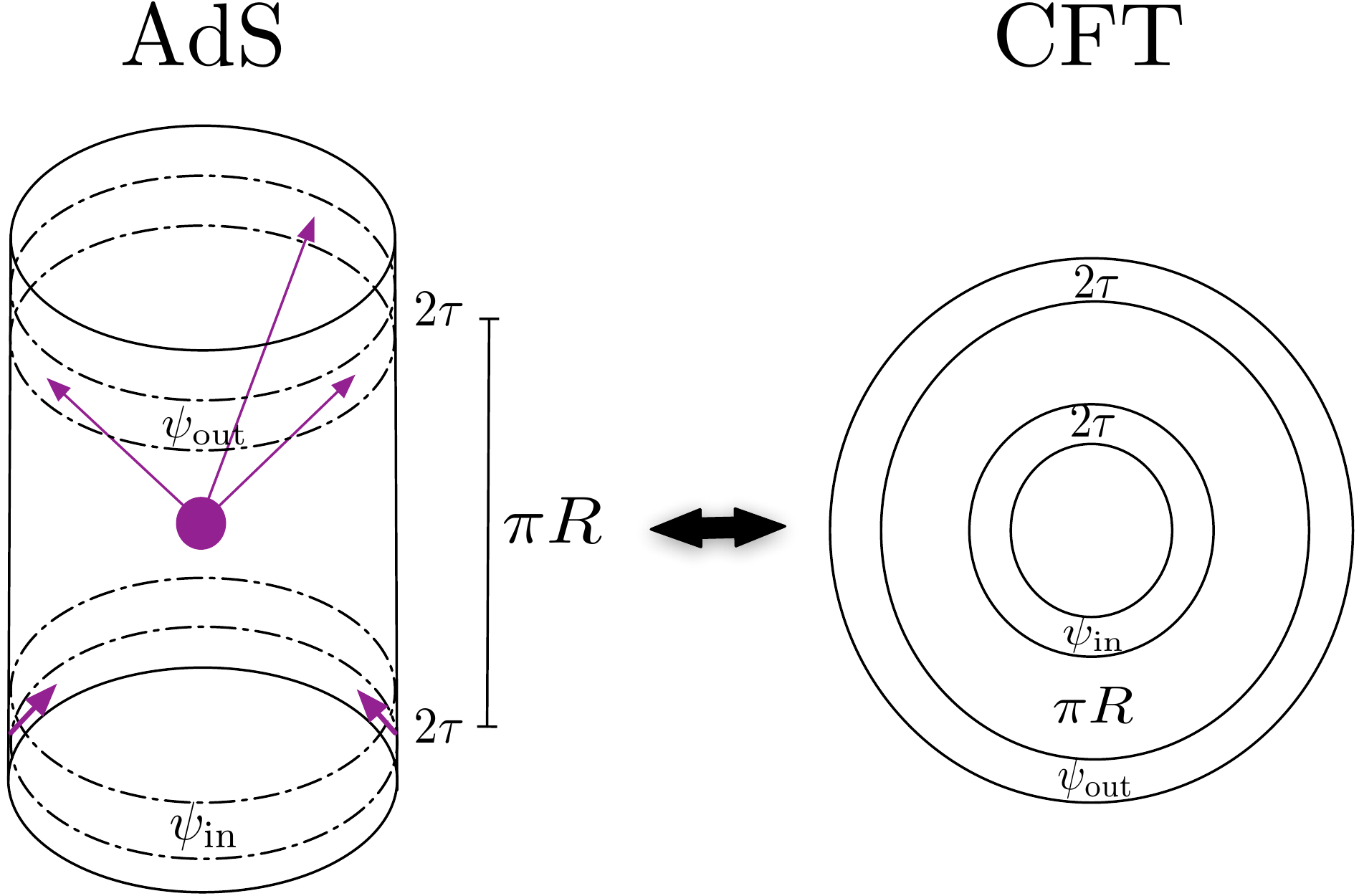}
\caption{ This figure shows how bulk scattering processes are setup in AdS/CFT.  
Creating the initial state involves smearing CFT operators over an annulus, which is just $S^{d-1} \times [-\tau, \tau]$ on the cylinder bounding AdS.  The integration over space and (dilatation) time in the CFT is necessary to select the direction and magnitude of the bulk momenta, respectively.  The regulator $\tau$ can be taken to infinity in the flat spacetime limit.  The final state is measured after a time $\pi R$, so that the particles have the opportunity to scatter exactly once.  
\label{fig:AdSCylinder}  }
\end{center}
\end{figure}

Now we will apply this procedure to the Mellin representation of the CFT correlator and derive equation (\ref{FlatSpaceLimitFormula}).  AdS in global coordinates translates into a Lorentzian radial quantization of the CFT, so the operator corresponding to the $i$th particle will be located at $x_i = e^{i t_i} \hat p_i$ in the $d$-dimensional spacetime where the CFT lives.  This means that up to a normalization factor coming from the single-particle states, the S-Matrix is given by the integral and limit
\be
T(s_{ij}) \propto \lim_{ \frac{R}{\tau}, \tau \to \infty} \int [ d \delta] \int_{- \tau \pm \frac{\pi R}{2} }^{\tau \pm \frac{\pi R}{2} } dt_i e^{i \omega_i t_i } M(\delta_{ij}) \prod_{i<j} \left(\cos \left( \frac{t_i - t_j}{R} \right) - \hat p_i \cdot \hat p_j + i \epsilon \right)^{-\delta_{ij}} \Gamma(\delta_{ij})
\ee
where the particles have energies $\omega_i$, and the measure of integration $[ d \delta ]$ is given in appendix \ref{app:MeasureOfIntegration}.  
Our task is to simplify and evaluate these integrals.  This will be possible because in the flat spacetime limit, it will turn out that the $t_i$ dependence is approximately Gaussian and that typically $\delta_{ij} \sim R^2$, so that the $\Gamma$ functions can be expanded in the Stirling approximation.  Then the $\delta_{ij}$ integrations can be performed in the stationary phase approximation, which will imply that $\delta_{ij} \propto s_{ij}$ and give a momentum conserving delta function.

Now let us begin to evaluate the integrals.  First note that either $|t_i - t_j| \ll R$, if both particles $i$ and $j$ are initial or final,  or else $| t_i - t_j - \pi R | \ll R$, if the $i$th particle is in the final state and the $j$th is in the initial state.  This means that we can approximate the cosine by its Taylor expansion.  But now the only difference between initial and final states is the sign of their momenta, so we can drop the distinction between initial and final states in the time integrals.  This just reproduces the familiar fact that we can assign all of the particles in a scattering amplitude to the in-state via analytic continuation.  Furthermore, we must anticipate that the flat-space S-matrix will contain a momentum-conserving $\delta$-function.  As discussed in \cite{Fitzpatrick:2011jn}, this is an immediate consequence of the spacetime symmetries -- since the conformal algebra reduces to the Poincar\'e algebra in the flat space limit, translations will be a good symmetry in that limit, and momentum will be conserved.  We will not {\it a priori} demand that the total momentum vanish; for convenience, then, let the center of mass momentum be $\vec{q}\equiv \frac{1}{n} \sum_i \vec{p}_i$, and introduce shifted momenta $\vec{p^\prime}_i\equiv \vec{p}_i - \vec{q}$ that do conserve momentum.  Since $\vec{q}$ will eventually be constrained to vanish in the flat-space limit, we will consider it to be parametrically smaller than the $p_i$'s, which grow $\propto R$.  

Expanding the $(t_i - t_j)$ dependent factors around the initial and final times $\pm \frac{\pi R}{2}$ simplifies the integrand, giving
\be
\int [ d \delta] \int_{-\tau }^{\tau} dt_i   e^{i \omega_i  t_i} M(\delta_{ij}) \prod_{i<j} \Gamma(\delta_{ij}) \left( \frac{s_{ij}}{\omega_i \omega_j} - \frac{t_{ij}^2}{R^2}  \right)^{- \delta_{ij}},
\ee
where we have defined $t_{ij} = t_i - t_j$.  Note that the overall time $\sum_i t_i$ only appears in the exponent, so one could immediately integrate over this direction to give an energy conserving delta function, removing a single time integration variable.
Now we will re-parameterize the $\delta_{ij}$ variables in a way that accords with the structure of the kinematics and the integral.  The momenta $p_i$ do not necessarily obey momentum conservation, so we can define
\ba
s'_{ij} &=& s_{ij} - \frac{2 n}{n-2} q \cdot (p_i + p_j) + \frac{2 n^2}{(n-1)(n-2) } q^2 ,
\ea
so that $\sum_{j \ne i} s'_{ij} =0$.  Now the $s_{ij}'$ are appropriate for on-shell states with momentum conservation.  Let us take
\be \label{eq:ChangeVariables}
\delta_{ij} = -\frac{R^2}{4 \alpha} \left( s'_{ij} + \epsilon_{ij} \right) + \Delta_{ij},
\ee
where we constrain 
\be \label{eq:EpsilonConstraints}
\forall \  i, \ \ \sum_{ j \ne  i}\epsilon_{ij}  = 0   \ \ \ \mathrm{and} \ \ \ \sum_{j\ne i}\Delta_{ij} = \Delta_i ,
\ee
so that the $\delta_{ij}$ satisfy the usual constraints and the $\alpha$ parameter deforms the $\delta_{ij}$ in a direction orthogonal to the $\epsilon_{ij}$.  This means that between $\alpha$ and the $\epsilon_{ij}$,  there are $n(n-3)/2$ free variables, as expected for the $\delta_{ij}$.  There is of course a Jacobian factor from this change of integration variables which includes a factor of $\alpha^{-\frac{n(n-3)}{2}-1}$, and also a single factor of $s_{ij}'$ due to the $\alpha$ direction.
We will perform the $\epsilon_{ij}$ integrations by saddle point, and we will see shortly that the saddle point is at $\epsilon_{ij}=0$ in
the large $R$ limit, so that only the $\alpha$ parameter remains as an integration variable.  

Next, let us expand the $\Gamma$ functions and $s_{ij}$ exponent at large $\delta_{ij}$:
\ba
&& \prod_{i< j} \Gamma(\delta_{ij})\left(\frac{s_{ij}}{\omega_i \omega_j} - \frac{t_{ij}^2}{R^2}   \right)^{- \delta_{ij}}   \nn\\
&&  \ \ \ \ \ \ \ \ \approx  \left( \frac{(4 \pi )(4 \alpha)}{ R^2} \right)^{\frac{n(n-1)}{4}} \left( \frac{ R^2}{4\alpha} \right)^{\Delta_\Sigma}
\left[ \prod_{i<j} \left( \frac{( s'_{ij} + \epsilon_{ij} ) \omega_i \omega_j }{s_{ij} } \right)^{\Delta_{ij}} \right]
\left[ \prod_{i<j} \left( s'_{ij} + \epsilon_{ij} \right)^{-\frac{1}{2}} \right] \nn\\
&& \ \ \ \ \ \ \ \ \times \exp \left( -\sum_{i<j} \frac{R^2}{4 \alpha} \left( s'_{ij} + \epsilon_{ij} \right) \log \left( \frac{ s'_{ij} + \epsilon_{ij} }
{\frac{s_{ij}}{\omega_i \omega_j} - \frac{t_{ij}^2}{R^2} } \right) \right)
\label{eq:expanded}
\ea
In what follows we will assume that these factors dominate the integrand in the flat space limit.  This assumption could be violated if the Mellin amplitude grows exponentially or faster at large $\delta_{ij}$.  Such growth would also call into question the convergence of the Mellin representation for the CFT correlator itself, but we cannot definitely rule it out.  Including the Jacobian from the change of $\delta_{ij}$ variables, the prefactor becomes
\ba \label{eq:Prefactor}
\frac{1}{\alpha}\left( \frac{R^2}{4\alpha} \right)^{\frac{n(n-3)}{2}}  (R^2)^{\Delta_\Sigma - \frac{n(n-1)}{4}} (4\pi)^{\frac{n(n-1)}{4}} (4\alpha)^{\frac{n(n-1)}{4} -\Delta_\Sigma} 
\left[ \prod_i \omega_i^{\Delta_i} \right] \left[ \prod_{i<j} s'^{-\frac{1}{2}}_{ij} \right] \delta_\tau \left( \sum_i \omega_i \right)
\ea
We see that the $\alpha^{-\Delta_\Sigma}$ factor has appeared as needed to reproduce the identical factor in equation (\ref{FlatSpaceLimitFormula}), while the other factors must cancel against the normalizations of the single particle states in equation (\ref{eq:SingleParticleNorm}) and the terms from the integrals below.
The argument of the exponential can be expanded and greatly simplified to give
\be
\exp \left[ i t_\omega + \frac{t_\omega^2}{4\alpha} - \frac{R^2}{4\alpha}\left( -(nq)^2 + \sum_{i<j} \frac{u_{ij}^2}{2 s'_{ij}} \right) \right]
\label{eq:gaussianeq}
\ee
where we have defined $t_\omega = \sum_i t_i \omega_i $ and
\ba
u_{ij} &=&  \epsilon_{ij} - \frac{2n}{n-2} q \cdot (p_i + p_j) + \frac{t_{ij}^2 \omega_i \omega_j}{R^2} .
\ea

We can now see equation (\ref{FlatSpaceLimitFormula}) beginning to take shape.  Since $nq$ is just the sum of the momenta, at large $R$ the Gaussian factor becomes a momentum conserving delta function multiplied by $\alpha^h$.   Furthermore, performing the Gaussian integral over $t_\omega$ would provide the crucial $e^{\alpha}$ factor present in equation (\ref{FlatSpaceLimitFormula}).  Finally, to complete the derivation we must integrate over $u_{ij}$ subject to the constraints on $\epsilon_{ij}$ from equation (\ref{eq:EpsilonConstraints}).  If we were to ignore the constraints, this computation would be trivial, but introducing Lagrange multipliers to incorporate the constraints and fix the range of integration over the time variables $t_i$ requires several more straightforward but lengthy computations, which we have relegated to appendix \ref{app:GaussianIntegrals}.  Schematically, starting from (\ref{eq:gaussianeq}), the constrained $u_{ij}$ integrations cancel $-1/2+n(n-3)/4$ factors of $\alpha$ and $R^2$ from the prefactor.  The subsequent integration over all time coordinates produces the energy-momentum conserving delta function and an additional $\alpha^{h-\frac{n}{2}+\frac{3}{2}}$, canceling the remaining factors of $s_{ij}$ and $\omega_i$.  This gives the final result
\be
T(s_{ij}) = \lim_{R \to \infty}\frac{1}{\NN} \delta^{d+1} \left( \sum_i p_i \right)  \int_{-i\infty}^{i\infty} d \alpha \ \!  e^{\alpha}  \alpha^{h - \Delta_\Sigma}  
M \left( \delta_{ij} = -\frac{R^2 s_{ij}}{4 \alpha} \right)
\ee
for the connected part of the S-Matrix, as desired.

\section{From Meromorphy to Analyticity}
\label{sec:Examples}

Our goal in this section will be to understand how the familiar analyticity properties of the S-Matrix follow from the flat space limit of the Mellin amplitude.  Specifically, we want to investigate branch cuts corresponding to the emission of multi-particle states and poles corresponding to finite-lifetime resonances.  On very general grounds, the Mellin amplitude is expected \cite{Mack, MackSummary, Fitzpatrick:2011ia} to have only simple poles on the real axis, so the more intricate analytic structure of the S-Matrix must emerge from the coalescence and resummation of the poles of the Mellin amplitude.

\subsection{AdS Exchanges}
\label{sec:AdSExchanges}

Let us begin by computing the flat space limit of the Mellin amplitude corresponding to the exchange of a massive field in the bulk of AdS.  This result will be very useful because in the next section we will express loop amplitudes as sums over bulk exchanges, via a procedure reminiscent of the K\"allen-Lehmann representation \cite{Dusedau:1985ue}.

To begin with, we can write the Mellin amplitude for an exchange as a sum 
\ba
M(\delta_{ij}) &=& \sum_m \frac{R_m}{\delta - (\Delta_5 + 2m)} ,
\ea
where for a 2-to-2 $s$-channel exchange, $\delta =  \Delta_1 + \Delta_2  - 2 \delta_{12}$ and $\Delta_5$ is the dimension of the operator dual to the bulk field being exchanged.  The residue was first computed in \cite{Penedones:2010ue} and can be easily obtained from the diagrammatic rules \cite{Fitzpatrick:2011ia, Paulos:2011ie}, it is
\be
R_m = -R^{5-2h}\frac{\Gamma(\Delta_{125,} - h)\Gamma(\Delta_{345,}-h)}{(4\pi^h)^3 \prod_{i=1}^4 \Gamma(\Delta_i - h + 1)} \times \frac{(1 - \Delta_{12,5})_m (1 - \Delta_{34,5})_m}{ m!  \Gamma(\Delta_5 - h + 1 + m) },
\label{eq:residues}
\ee
where we will often use the notation $2 \Delta_{a_1...a_k,b_1...b_l} = \Sigma_i^k \Delta_{a_i} - \Sigma_j^l \Delta_{b_j}$.
In the flat space limit, the bulk energies dual to CFT dimensions must be taken to be very large compared to the curvature scale $1/R$ in AdS, so that in particular as $R \to \infty$ we must have $\Delta_5 \propto R$ and $\delta \propto s R^2$ for a massive exchange.  This means that $R_m$ is dominated by large values of $m$ so that $m \sim \Delta_5^2 \propto R^2$, and we find that
\ba
R_m &=& C \left( m^{h - \Delta_{1234,}} e^{-\frac{\Delta_5^2}{4m}} \right) ,
\ea
where the proportionality constant $C$ is
\ba
C= -\frac{R^2}{2}  \left( \frac{\Delta_5^2}{4} \right)^{\Delta_{1234}-h-1}\NN .
\ea
and $\NN$ is the normalization factor from equation (\ref{FlatSpaceLimitFormula}).
 Now, when we take the flat space limit of the Mellin amplitude, we can replace the sum over $m$ with an integral,  giving
\be
T(s) = \frac{C}{\NN}\int_{-i \infty}^{i \infty} \frac{d \alpha}{2 \pi i} \alpha^{h - \Delta_{1234}} e^{\alpha} \int_0^\infty dm \frac{ m^{h - \Delta_{1234}} e^{-\frac{\Delta_5^2}{4m}}  }{ \frac{R^2 s}{2\alpha} - (\Delta_5+2m) },
\ee
where the integration contour in $\alpha$ runs to the right of all the poles and the branch cut (which arises due to the irrational power in the definition of the flat space limit), and $2 \Delta_{1234} = \Delta_1 + \Delta_2 + \Delta_3 + \Delta_4$.  To compute these integrals, we can deform the $\alpha$ contour to pick up a contribution from the poles and a contribution from the discontinuity across the branch cut.  Changing integration variables from $m$ to $x = 1/m$ for convenience, the pole contribution is
\ba
\frac{1}{2} \int dx \ \! e^{-x \frac{s-\Delta_5^2}{4}} \left(-\frac{s R^2}{4} \right)^{1+h-\Delta_{1234}} = \frac{2}{s R^2 -\Delta_5^2} \left(-\frac{s R^2}{4} \right)^{1+h-\Delta_{1234}} 
\ea
We see that the result has a pole in the correct place, but there is an unusual-looking overall $s$-dependent factor.
The contribution from the branch cut is
\ba 
 \frac{\sin (\pi (1+h-\Delta_{1234}))}{\pi} \int_0^\infty d \alpha \int_0^\infty dx e^{-\alpha} \left( \frac{\alpha}{x} \right)^{1+h-\Delta_{1234}} \frac{2 e^{-\frac{x \Delta_5^2}{4} }}{-R^2 s x + 4 \alpha} .
\ea
It is easy to evaluate these integrals by changing integration variables from $x$ to $y=x/\alpha$, which makes the $d \alpha$ integration and subsequent $dy$ integration straightforward, giving
\ba
  \frac{2}{s R^2+\Delta_5^2} \left( - \left( - \frac{sR^2}{4} \right)^{1+h - \Delta_{1234}} + \left( \frac{\Delta_5^2}{4} \right)^{1+h-\Delta_{1234}} \right)
\ea
Thus, the branch cut piece combines with the pole piece to cancel off the strange $s$-dependence, and incorporating the overall coefficient leaves us with the expected flat space propagator
\be
T(s) = \frac{1}{s - m_5^2}
\ee
where $m_5 \equiv \Delta_5/R$.  

\subsection{Computing Loop Diagrams a la K\"allen-Lehmann}

In flat spacetime, the two point functions of local operators can always be decomposed in the K\"allen-Lehmann representation \cite{Weinberg:1995mt} as
\be
\langle \Phi(k) \Phi(-k) \rangle_{\mathrm{flat}} = \int_0^\infty d\mu \frac{\rho(\mu^2)}{k^2 + \mu^2 + i \epsilon}
\ee
where the function $\rho(\mu)$ is positive and real.  This decomposition follows from unitarity and Lorentz invariance \cite{Weinberg:1995mt}, and it expresses the two point function of general operators as a sum over the two point functions of free fields with different masses.  Now we will compute some simple loop amplitudes by using the AdS analog of this representation, which was first discussed in \cite{Dusedau:1985ue} long before the discovery of AdS/CFT. 

\begin{figure}[t!]
\begin{center}
\includegraphics[width=0.85\textwidth]{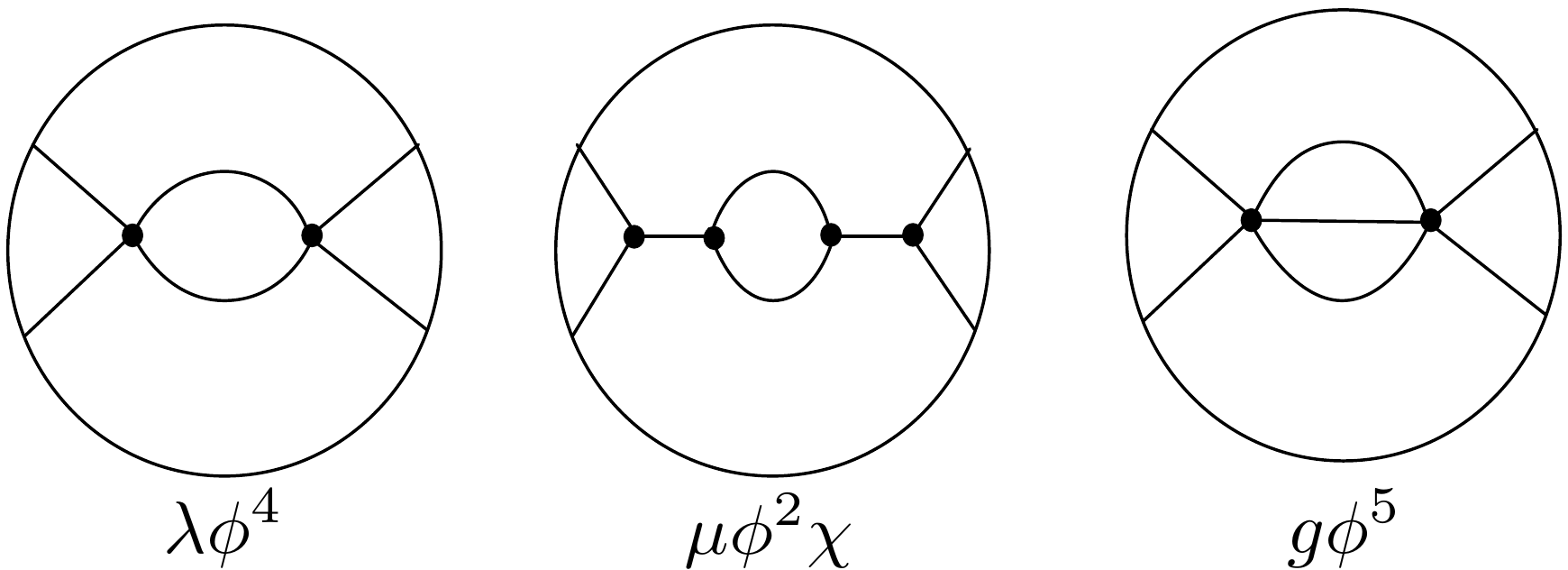}
\caption{ This figure shows the 1-loop and 2-loop diagrams that we have computed using the Mellin space version of the K\"allen-Lehmann representation.  Further generalizations are straightforward.
\label{fig:LoopDiagrams}  }
\end{center}
\end{figure}

Consider computing the standard 1-loop bubble diagram from $\lambda \phi^4$ theory in AdS, which is pictured in figure \ref{fig:LoopDiagrams}.  In position space, this would require us  to calculate
\be
\int d^{d+1}X d^{d+1} Y G_{\Delta_1}(P_1, X) G_{\Delta_2}(P_2, X) G_\Delta(X,Y)^2 G_{\Delta_3}(Y, P_3) G_{\Delta_4}(Y, P_4)
\ee
The only difference between this calculation and the computation of a tree level exchange in $\phi^2 \chi$ is the replacement of the single propagator $G_{\Delta_\chi} (X,Y)$ with the propagator squared $G_\Delta(X,Y)^2$.  But the square of the bulk-to-bulk propagator is just the two point function of the local bulk field $\phi^2$, so that
\be \label{eqn:1LoopMellin}
G_\Delta(X,Y)^2 = \left\langle \phi^2(X) \phi^2(Y)  \right\rangle
\ee
Let us compute this two-point function with a Mellin space K\"allen-Lehmann representation.  This means that we will write
\be \label{eqn:PropSquared}
G_\Delta(X,Y)^2 = \sum_n N_{2,\Delta}(n) G_{2 \Delta + 2n}(X,Y)
\ee
Using the diagrammatic rules \cite{Fitzpatrick:2011ia, Paulos:2011ie} from section \ref{sec:DiagrammaticRules}  it becomes trivial to compute the loop diagram once we know $N_{2,\Delta}$.  One can compute the function $N_{2,\Delta}(n)$ in Mellin-space, but a direct derivation via an inner product, as was used in \cite{Katz}, is probably the simplest method.   In appendix \ref{app:KLDerivation} we show that for any two scalar propagators
\ba \label{eqn:ProductOfPropagators}
G_{\Delta_1}(X,Y) G_{\Delta_2}(X,Y) &=& \sum_n  a_{\Delta_1, \Delta_2}(n) G_{\Delta_1 + \Delta_2+2n}(X,Y),  \ \ \mathrm{where} \\
a_{\Delta_1,\Delta_2}(n) &=&  \frac{(h)_n}{2 \pi^h n! }  \frac{  (\Delta_1 + \Delta_2 + 2n)_{1-h}  (\Delta_1 + \Delta_2 + n -2h +1)_n }
{(\Delta_1 + n)_{1-h} (\Delta_2 + n)_{1-h} (\Delta_1 + \Delta_2 + n -h)_n }. \nonumber
\ea
and recall that the Pochhammer symbol $(x)_y = \Gamma(x+y)/\Gamma(x)$, and $2h = d$, the spacetime dimension of the CFT.
It is useful to note that for $\Delta \ll n$, as required for the flat space limit, we have
\be \label{eqn:FSLPropSquared}
N_{2,\Delta} (n) = a_{\Delta, \Delta}(n) \approx
\frac{2}{(4 \pi )^{h} \Gamma (h) } 
n^{2(h-1)}.
\ee
This means that in this limit $N_{2,\Delta}(n)$ has a simple power law dependence on the parameter $n$ which determines the energy of the bulk states being exchanged.

There is no need to stop at 1-loop\footnote{We apologize to readers uninterested in juvenile computational showmanship.}.  For example, in $g \phi^5$ theory at 2-loops there is a 4-pt bubble diagram involving three propagators, as pictured in figure \ref{fig:LoopDiagrams}, which we can easily compute.  We see that
\be
G_\Delta(X,Y)^3 = \sum_n N_{3,\Delta}(n) G_{3 \Delta + 2n}(X,Y),
\ee
where we have that
\be
N_{3,\Delta}(n) = \sum_{m=0}^n a_{\Delta, \Delta}(m) a_{\Delta, 2\Delta+2m}(n-m) .
\ee
It is a non-trivial task to perform the sum explicitly.  However, it is easy to take the flat space limit of the sum:
\ba
N_{3,\Delta}(n)&\approx& \int_0^n  dm  \frac{4 (n^2 - m^2)^{2(h-1)} }{(4\pi)^{2h} \Gamma^2(h)} = \frac{n^{4h-3}}{2 (2\pi)^{2h} (h-\frac{1}{2})_h \Gamma(h)}.
\ea
This has precisely the correct scaling in $n$ for the K\"allen-Lehmann representation of the 2-loop bubble diagram of figure \ref{fig:LoopDiagrams} in $2h+1$ dimensions.

\subsection{Branch Cuts and Resonances}

\begin{figure}[t!]
\begin{center}
\includegraphics[width=0.75\textwidth]{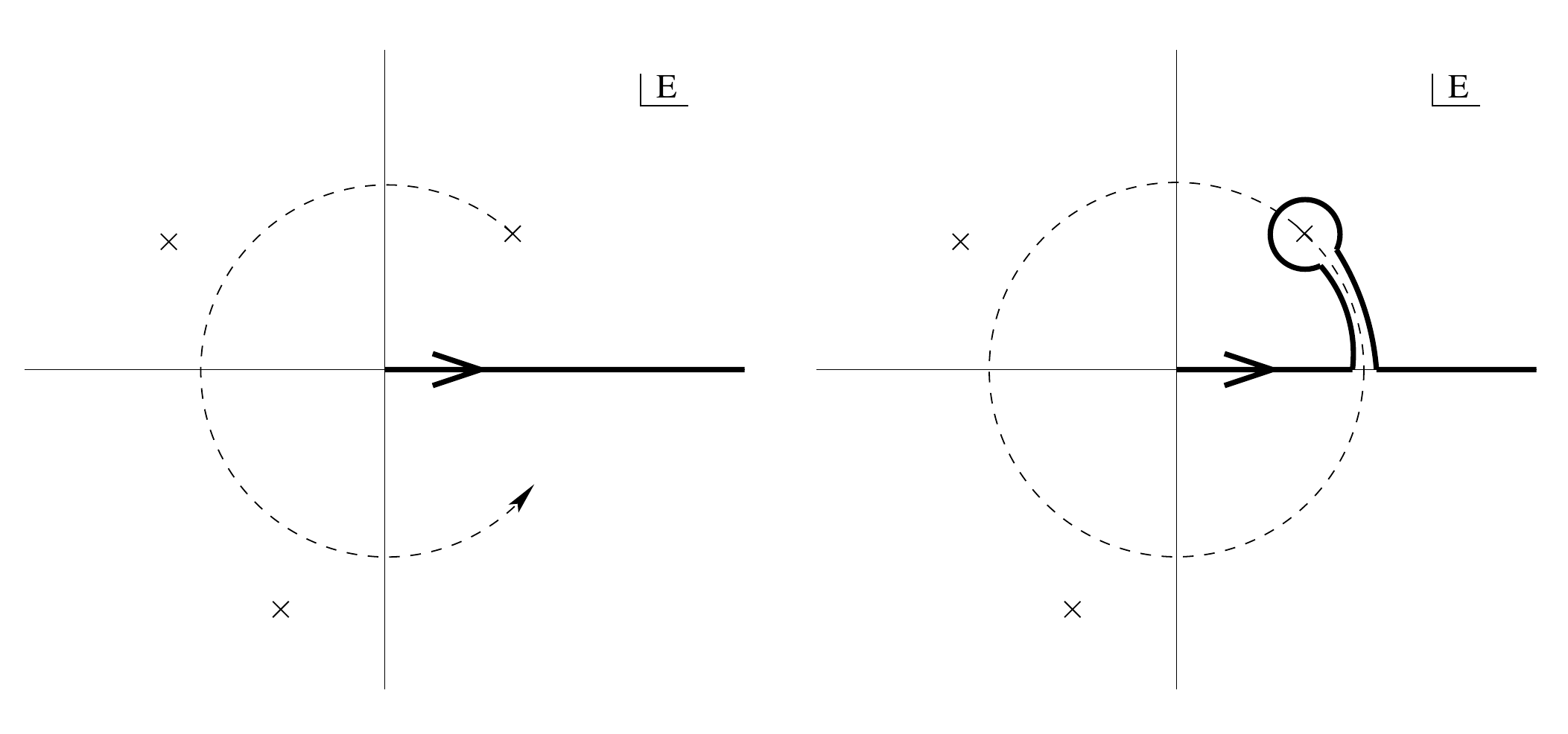}
\caption{ This figure shows how poles in the integrand of a contour integral lead to branch cuts in the result, due to the necessity of deforming the contour as the pole moves.
\label{fig:BranchCutsfromPoles}  }
\end{center}
\end{figure}

Now let us see how two familiar non-analytic features of scattering amplitudes, namely branch cuts from multi-particle intermediate states and finite-width resonances from unstable particles, arise from the Mellin amplitude.

These phenomena are both connected with the flat spacetime limit, where the AdS length $R \to \infty$. However, they do not depend on the intricacies of AdS/CFT, but are instead generic consequences of defining a theory inside a finite `box' and then taking the size of the box to infinity.  AdS acts as an IR-regulating cavity, and in the flat spacetime limit the cavity's volume becomes infinite, so that the discrete spectrum of modes approaches a continuum.  Branch cuts can arise in this limit when a line of poles coalesces, because the separation between poles and the residues of each pole are both proportional to $1/R$.  The sum over the discrete spectrum becomes indistinguishable from an integral, and as shown in figure \ref{fig:BranchCutsfromPoles} such integrals develop branch cuts due to the poles of their integrands.  Re-summing propagators with branch cuts, as is necessary for the central diagram of figure \ref{fig:LoopDiagrams} for $\mu \phi^2 \chi$ theory, leads to finite-width $\chi$ particles.  Thus the Mellin amplitude, whose poles all lie on the real axis, can produce poles off the real axis in the S-Matrix.

Let us first see how branch cuts arise at 1-loop in $\frac{\lambda}{24} \phi^4$ theory, for the first diagram in figure \ref{fig:LoopDiagrams}.  For this purpose we need to take the flat space limit of equation (\ref{eqn:PropSquared}).
We saw in section \ref{sec:AdSExchanges} that an AdS propagator $G_{\Delta } \to \frac{1}{s - (\Delta)^2}$ in the flat space limit, so we can directly use the function we derived in equation (\ref{eqn:ProductOfPropagators}) to find the 1-loop scattering amplitude in $\frac{\lambda}{24} \phi^4$ theory.    In the flat spacetime limit the sum over $n$ becomes an integral, and so we can use the large $n$ approximation in equation (\ref{eqn:FSLPropSquared}),  giving 
\be \label{eqn:1LoopFlatSpacePhi4}
\mathcal{M}^{\mathrm{1-loop}}(s) = \frac{2 \lambda^2 }{(4 \pi )^{h} \Gamma (h) } \int_0^\infty dn \frac{n^{2(h-1)}}{s - (2 \Delta + 2n)^2}
\ee
Note that this is precisely of the form expected from the K\"allen-Lehmann representation of the bulk theory in flat spacetime, and it agrees with the usual result for $\frac{\lambda}{24} \phi^4$ theory in $2h+1$ dimensions.  Similarly, the 2-loop result in $g \phi^5$ theory is 
\be 
\label{eqn:2LoopFlatSpacePhi5}
\mathcal{M}^{\mathrm{2-loop}}(s) \propto \int_0^\infty dn \frac{n^{4h-3}}{s - (3 \Delta + 2n)^2}
\ee
in the flat space limit, again in the K\"allen-Lehmann representation.  This has the correct scaling for this theory in $2h+1$ bulk dimensions.

From equations (\ref{eqn:1LoopFlatSpacePhi4}) and (\ref{eqn:2LoopFlatSpacePhi5}) we can immediately see why the loop amplitudes have branch cuts. 
As pictured in figure \ref{fig:BranchCutsfromPoles}, when a pole of the integrand moves across the contour of integration, the contour must be deformed around the pole.  If the poles makes a complete circle about an endpoint of the contour, then the contour can return to its original location, but it must also include a tiny circle about the pole.  Thus the residue of the integrand's pole becomes the discontinuity across the branch cut of the integral. 

\begin{figure}[t!]
\begin{center}
\includegraphics[width=0.85\textwidth]{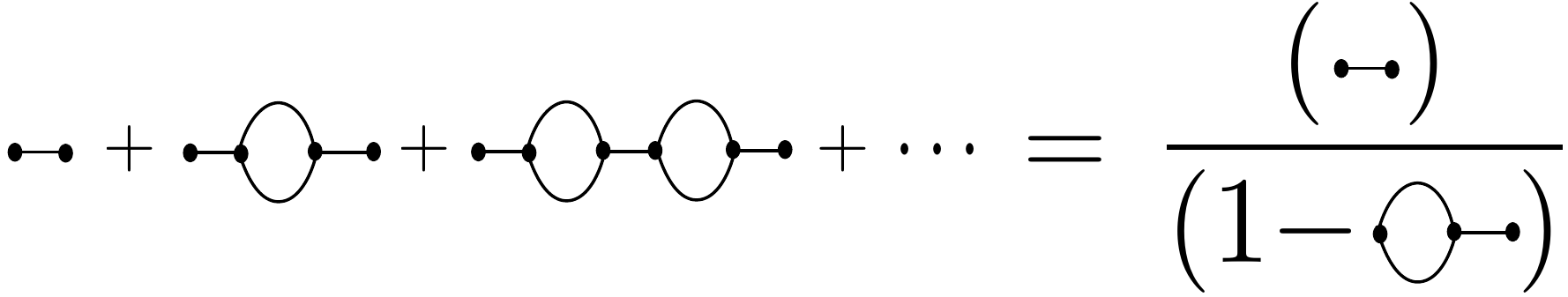}
\caption{ This figure shows the standard picture of how 1-PI diagrams are resummed to shift propagator poles.  This familiar momentum-space process also works in Mellin space.
\label{fig:PropagatorResummation}  }
\end{center}
\end{figure}

Now let us discuss how we obtain resonances in the flat space limit, beginning with the Mellin space version of the resummation of 1-PI diagrams, which is pictured in figure \ref{fig:PropagatorResummation}.  If we compute the resummed diagram corresponding to the one-loop 1-PI diagram in $\mu \phi^2 \chi$ theory, we find
\ba
M^{\mathrm{sum}} &=& \mu^2 \sum_{m_1,m_2} \frac{V_{12,\Delta_\chi}(m_1) S_{\Delta_\chi}(m_1) }{\delta - (\Delta_\chi + 2m_1)} \times \left[ \frac{1}{1-\Pi_{m_i, m_j}(\delta) } \right]_{m_1,m_2} \times
V_{\Delta_\chi,34}(m_2) 
\label{eq:imagpole1}\ea
where we have the infinite dimensional matrix in $m_i$ space
\ba
\Pi_{m_1, m_2}(\delta) &=& \mu^2\sum_{m_3} \left[ \sum_n \left( N_{2,\Delta_\phi}(n) \frac{V_{\Delta_\chi, 2\Delta_{\phi} +2n}(m_1, m_3) S_{2 \Delta_\phi + 2n}(m_3)   } {\delta - (2 \Delta_\phi + 2n + 2m_3)} \right) \right.  \nonumber \\
&& \left. \times \frac{V_{2\Delta_{\phi} +2n, \Delta_\chi}(m_3, m_2) S_{\Delta_\chi}(m_2)}{\delta - (\Delta_\chi + 2m_2)} \right] 
\label{eq:imagpole2}
\ea
We have written this result so that the first term is simply the answer for $\chi$-exchange at tree level. The first line of the expression for $\Pi$ gives the K\"allen-Lehmann-type representation for the 1-loop bubble diagram, and the second line provides the final $\chi$-propagator, with the last term replacing the final vertex in the tree-level $\chi$-exchange.  Since in position space in AdS we are connecting pairs of propagators, we must use 2-pt vertex functions which follow from the vertex rule in equation (\ref{eqn:Vertices}). 

We would like to see that $\Pi$ develops an imaginary piece from the loop in the flat space limit, where we take $\Delta_\chi = m_\chi R$ and $\delta = -\frac{R^2 s}{4 \alpha}$ as usual, but we leave $\Delta_\phi$ finite so that the $\phi$ particles become effectively massless.  In this limit, the Mellin amplitude is dominated by terms in the sums with $m_i \propto (R \Delta)^2$.  Rather than performing the sums in eq. (\ref{eq:imagpole1},\ref{eq:imagpole2}) directly, we will again take advantage of the fact that the bubble in the diagram can be written as a sum over single propagators corresponding to the double-trace operators, as in eq. (\ref{eqn:ProductOfPropagators}). Written this way, the bubble diagrams are identical in form to the mixing of $\chi$ with an infinite tower of states of dimensions $2\Delta_\phi + 2n$ for $n \geq 0$.  
The contribution from these states can be determined by diagonalizing the mass matrix
\ba
m_{\rm eff}^2 &=& \left( \begin{array}{ccccc} \Delta_\chi^2 & R^2 \lambda_{\rm eff}(0) &  R^2 \lambda_{\rm eff}(1) & R^2 \lambda_{\rm eff}(2) &  \cdots \\
R^2 \lambda_{\rm eff}(0) & (2\Delta_\phi)^2 & 0 & 0 & \cdots  \\
R^2 \lambda_{\rm eff}(1) & 0 & (2\Delta_\phi+2)^2  & 0 & \cdots \\
R^2 \lambda_{\rm eff}(2) & 0 & 0 & (2\Delta_\phi+4)^2  & \cdots \\
\vdots & \vdots & \vdots  & \vdots & \ddots   
\end{array} \right),
\label{eq:effmsq}
\ea
where 
\be 
\lambda_{\rm eff}(n) \equiv \lambda \sqrt{\frac{N_{2\Delta_\phi}(n)}{ R^{2h - 1}}}
\ee 
is the effective off-diagonal mass-mixing term, and we have approximated mass as dimension since we are interested in the flat-space limit.  The factor of $N_{2 \Delta_\phi}(n)$ accounts for the phase space of the 2-particle $\phi$ states, so that their contribution to the two point function can be understood via mixing.   One finds that with this coupling, the mixing amplitude also correctly computes the decay rate of $\chi \to 2 \phi$ particles.  Note that there is nothing particularly  `holographic' about this method; one would use identical techniques to understand how resonances arise when one restricts a quantum field theory to a finite sized box with a discrete spectrum.

We expect that the sum of all 1-PI diagrams must be given by the Mellin amplitude that would arise from adding a mass mixing term as in eq. (\ref{eq:effmsq}) above, but let us see how we could obtain this result directly by using the functional equation.  First, note that when we act with the functional equation on the Mellin amplitude $M$ for the resummation of 1PI diagrams, we just obtain $M$ back again, plus a constant for the reduction of the single propagator:
\ba
m_{eff}^2 M(\delta_{LR} )  = \left( -\delta_{LR} (\delta_{LR} - d) - \Delta (d-\Delta) \right) M(\delta_{LR})  - (\delta_{LR} - 2 \Delta)^2 M(\delta_{LR}-2) - 1
\ea
We have extended $M(\delta_{LR})$ to  a matrix whose first row and column correspond to $\chi$ propagation, so that only the $M_{11}$ element actually appears in the $\phi$ four-point amplitude, while the rest of the matrix involves mixing with two particle $\phi$ states.  Now, this clearly takes the form of a mass-mixing, and the matrix operation on $M(\delta_{LR})$ can be diagonalized. Let the $a$th eigenvalue of the mass matrix be $\Delta_a$, then we may write the corresponding Mellin amplitude as
\ba
M(\delta_{ij}) &=& \sum_{a} S_{1 a} D_a(\delta_{LR}) S^T_{a 1}, \nn\\
D_a (\delta_{LR}) &=& \sum_m \frac{R_m(\Delta_a)}{\delta_{LR} - (\Delta_a + 2m)}
\ea
where $R_m$ is the formula for the residues from eq. (\ref{eq:residues}) with dimension $\Delta_a$, and $S_{ij}$ is the matrix of eigenvectors of eq. (\ref{eq:effmsq}).  We already established in section \ref{sec:AdSExchanges} that the flat space limit of $D_a$ is simply a propagator with mass $\Delta_a$, so we can use these equations to compute the flat space limit of the sum of 1-PI propagators pictured in figure \ref{fig:PropagatorResummation}.

To see how the imaginary piece of the pole in the holographic S-matrix emerges, we need only study the eigen-decomposition of $m_{eff}^2$ in the flat-space limit with $s$ very near the real part of the pole, so that $s = \Delta_\chi^2 + \delta s$.  In this limit the mixing terms are dominated by double-trace operators with dimension near  $\Delta_\chi$, where `nearness'  is determined by the magnitude of $\lambda_{eff}$.  Replacing the sum on dimensions $\Delta_a$ with an integral over a parameter $\eta$ in the flat space limit, we see that for small $\delta s$ we can approximate
 \ba
  S_{1 a}^2(\eta) &\sim& \frac{\lambda_{eff} R}{m_\chi} \frac{ 1 }{  \eta^2 + \frac{R^2 \lambda_{eff}^2}{m_\chi^2} } .
 \ea
where we are evaluating $\lambda_{eff}$ at $n = R m_\chi$.  This simply means that the number of modes that mix significantly with $\chi$ is proportional to $R \lambda_{eff}/m_\chi$.  Considering the entries of $m_{eff}$ near $2 \Delta_\phi + 2n = \Delta_\chi$, one finds that in the continuum limit the eigenvalues can be approximated by $\delta s + \lambda_{eff} \eta$, so using the results of section \ref{sec:AdSExchanges} on the flat space limit of the propagator $D_{n_*}(\delta_{LR})$, we find that the sum of 1-PI diagrams can be approximated by
\ba
\int_{-\infty}^\infty d \eta \left(\frac{ \lambda_{eff} R }{m_\chi} \frac{ 1 }{  \eta^2 + \frac{R^2 \lambda_{eff}^2}{m_\chi^2} }  \right)
\frac{1 }{\delta s + \lambda_{eff}  \eta + i \epsilon}
 &\approx&
   \frac{i \pi}{ \delta s + i \lambda^2 m_{\chi}^{2h-3}} 
\ea
which shows the usual Breit-Wigner structure, with a resonance pole that has moved off of the real axis.  One can obtain a more precise result by diagonalizing the $m_{eff}^2$ matrix numerically.

\subsection{A Comment about Meromorphy and Locality}

An extremely interesting example of how bulk locality can be  rigorously derived from conditions on a CFT was provided by the analysis of \cite{JP, Heemskerk:2010ty}.  In that case, it was assumed that the spectrum of low-dimension operators in the CFT included only a finite number of single-trace scalar primaries and that $\frac{1}{N}$ corrections affected conformal blocks only up to a maximum spin $L$.  By imposing crossing symmetry on the CFT correlation functions and then counting the dimension of the space of possible solutions, the authors of \cite{JP} were able to show that all such CFTs have local AdS dual descriptions.  It is interesting to see how this result is reproduced in terms of analyticity of the holographic S-matrix.

The point is that, as was explained in \cite{Penedones:2010ue}, such solutions must have Mellin amplitudes that are polynomials in $\delta_{ij}$'s.  This is easy to see by noting that the presence of a pole in the Mellin amplitude will immediately require the inclusion of conformal blocks of arbitrarily large spin.  For example, a pole in $\delta_{12}$ in the Mellin amplitudes for a CFT 4-point correlator implies an  equivalent pole in $\delta_{13}$, due to crossing symmetry.  But the decomposition of the $\delta_{13}$ pole in the 12-34 channel (the `s channel') will contain arbitrarily high powers of $\delta_{13}$, implying the presence of arbitrarily large spins.  Thus a Mellin amplitude with only conformal blocks of bounded spin cannot have any poles, and similarly it cannot included exponential dependence on the $\delta_{ij}$, so it must be a polynomial.  However, we know that polynomial Mellin amplitudes turn into S-Matrices that are polynomials in the Mandelstam invariants $s_{ij}$, and that are therefore obviously analytic and exponentially bounded.  

\section{Conformal Blocks and Black Holes}
\label{sec:ConformalBlocksandBH}

The simple unitarity relation that follows from inserting $\bf 1$ as a sum over states
\be
A_4(x_i) = \sum_\alpha \langle \CO_1(x_1) \CO_2(x_2) |\alpha \rangle \langle \alpha | \CO_3(x_3) \CO_4(x_4) \rangle
\ee
provides an interesting method for representing CFT 4-pt functions, via the operator-state correspondence.  This follows because we can express any \emph{primary} state in the sum as $| \alpha_p \rangle = \CO_\alpha | 0 \rangle$ for some local primary operator $\CO_\alpha$, and then the properties of all of the descendent states will be fixed by conformal invariance.  The functional form contributed to a 4-pt function by the exchange of a primary state/operator of dimension $\Delta$ and angular momentum $\ell$ along with all of its descendants is called a conformal block \cite{Belavin:1984vu, Dolan:2000ut, Dolan:2003hv, Costa:2011dw}.  We can write any CFT 4-pt function or 4-pt Mellin amplitude as a sum over conformal blocks 
\be \label{eqn:ConformalBlockDecomp}
M_4(\delta_{ij}) = \sum_\alpha C(\Delta_\alpha, \ell) B_{\Delta_\alpha, \ell}(\delta_{ij})
\ee
where the coefficients $C(\Delta_\alpha, \ell)$ are determined by the magnitude of the 3-pt functions of primary operators.  For a very readable discussion of conformal blocks and their applications in constraining the properties of CFTs, see \cite{Rattazzi:2008pe}.

We will make more extensive and essential use of conformal blocks in  our forthcoming work on the unitarity of the S-Matrix \cite{Unitarity}, but for now we provide these objects as an example of interesting functions on Mellin space that are not strictly analytic in the flat space limit.  Since conformal blocks correspond to the exchange of an operator with definite angular momentum and definite dimension, in the flat space limit they must also have definite angular momentum and definite energy.  This suggests that the flat space limit of a conformal block should be
\be
B_{\Delta,\ell}(\delta_{ij}) \to P_\ell(\cos \theta) \delta( s - \Delta^2)
\ee
where $P_\ell$ is just the appropriate Legendre or Gegenbauer polynomial.  We will derive this fact in the next section, but note that it gives a direct connection between the Bootstrap program for solving CFTs and the S-Matrix program for determining scattering amplitudes.  It suggests that the Bootstrap program, even for special theories such as $\mathcal{N} = 4$ SYM, should be at least as challenging as using analyticity and unitarity to directly construct the full non-perturbative S-Matrix of superstring theory.

In the flat space limit, the conformal block decomposition turns into a trivial integral over CFT dimensions, and we find that the partial wave decomposition of the 2-to-2 scattering amplitude must simply be
\be
\mathcal{S}_\ell(s) = C(\sqrt{s}, \ell)
\ee
where we have used the symbol $\mathcal{S}$ to emphasize that this is the full scattering amplitude, including the `$\bf 1$' piece.  This means that we can use known properties of bulk scattering amplitudes to constrain the conformal block decomposition of dual CFTs.  In section \ref{sec:BH} we will use the expected form of the trans-Planckian scattering amplitudes corresponding to Hawking evaporation to make a general comment about the conformal block decomposition in CFTs with AdS duals that are well-described by effective field theory.

\subsection{Flat Space Limit of a Conformal Block}

Now we will directly compute the flat space limit of the conformal blocks, which we review in Mellin space in appendix \ref{sec:ConformalBlocks}.  As Mack \cite{Mack} showed, in Mellin space the conformal blocks of angular momentum $\ell$ and dimension $\Delta$ in the $s$-channel have only a polynomial dependence on the $\delta_{13}$ and $\delta_{14}$ variables analogous to the mandelstam invariants $t$ and $u$.  Although these Mack polynomials appear somewhat complicated, they simply encode the fact that the conformal blocks have a fixed angular momentum, and in the flat space limit they reduce to Legendre or Gegenbauer polynomials in $\cos \theta$.  We show these straightforward but technical facts in appendix \ref{sec:ConformalBlocks}. 

The Mack polynomials multiply a universal function that encodes a more interesting dependence on $\delta_{12}$, analgous to the mandelstam invariant $s$, and $\Delta$, the dimension of the conformal block.  This function takes the form\footnote{We thank J. Penedones, S. Raju, and B. van Rees for discussions and collaboration on conformal blocks in the Mellin representation, and in particular for collaboration leading to the discovery of the factor in parentheses in this equation.

It should also be noted that we are using a different normalization convention from most of the recent literature on conformal blocks.  For instance, to convert to the normalization used in \cite{JP}, with $\Delta_1= \Delta_2 $ and $\Delta_3 = \Delta_4$, one should take eq.~(\ref{eq:scalarconfblocks}) and multiply by
\be
N_{\Delta}^\ell= \frac{i \pi  2^{-l-1} \csc ^2\left(\frac{1}{2} \pi  (\Delta +2 h)\right)
   \Gamma (-h+\Delta  +1) \Gamma (l+\Delta ) (-2 h+\Delta 
   +2)_{-l}}{\Gamma \left(\frac{l}{2}+\frac{\Delta }{2}\right)^2 \Gamma
   \left(\frac{1}{2} \left(\Delta  -2
   \left(h+\frac{l}{2}-1\right)\right)\right)^2},
   \ee
and use $N_{\Delta}^\ell  B_{\Delta}^\ell(\delta)$.  
 }
\be
B_\Delta^\ell(\delta) = e^{ \pi i(h- \Delta)} \left( e^{i \pi (\delta+ \Delta + \ell -2h)} -1\right)  \frac{\Gamma \left(\frac{\Delta-\ell-\delta}{2} \right) \Gamma \left(\frac{2h-\Delta - \ell - \delta}{2} \right) }{ \Gamma \left(\Delta_a-\frac{\delta}{2} \right) \
 \Gamma \left(\Delta_b - \frac{\delta}{2} \right) } 
 \label{eq:scalarconfblocks}
\ee
The factor in parentheses cancels the `shadow' poles associated with the second $\Gamma$ function in the numerator, so that we are left with only physical  poles associated to a single primary operator.  The normalization factor has been chosen to ensure that the residues of the physical poles are real.  The variables $2\Delta_a \equiv \Delta_1 + \Delta_2$ and $2\Delta_b \equiv \Delta_3 + \Delta_4$ are defined for convenience.  

When we take the limit $R \to \infty$, we must take $\Delta = MR$ while fixing the angular momentum $\ell$ to be a finite integer, so that the conformal block will have a non-zero bulk energy $M$ in the flat space limit.  Note that in this limit there is no difference between the dimension and twist of the block.  Plugging $B_\Delta^\ell(\delta)$ this into our formula for the flat space limit gives
\be
\lim_{R \to \infty} \int_{- i \infty}^{i \infty} \frac{d \alpha}{2 \pi i} e^\alpha \alpha^{h - \Delta_{1234}} e^{ \pi i(h- \Delta)} \left( e^{\pi i \frac{R^2s}{2\alpha} } -1 \right) \frac{\Gamma \left(\frac{R^2 s }{2\alpha} - \Delta_{12,5} \right) \Gamma \left(\frac{R^2 s }{2\alpha} - \tilde\Delta_{12, 5} \right) }{\Gamma \left( \frac{R^2 s}{2\alpha} \right) \Gamma \left( \frac{R^2 s}{2\alpha} - \Delta_{12,34} \right) }
\ee
where $2 \Delta_{12,5} = \Delta_1 + \Delta_2 - \Delta - \ell$ and $2 \tilde \Delta_{12, 5} = \Delta_1 + \Delta_2 - 2h - \ell+ \Delta$.  
Since we are taking the AdS scale to infinity, we can expand the $\Gamma$ functions using Stirling's approximation, giving
\ba
\lim_{R \to \infty} \int_{- i \infty}^{i \infty} \frac{d \alpha}{2 \pi i} e^\alpha \alpha^{h - \Delta_{1234}} e^{ \pi i(h- \Delta)} \left( e^{\pi i \frac{R^2s}{2\alpha} } -1 \right) 
\left( \frac{R^2 s}{2 \alpha} \right)^{\Delta_{1234} - h - \ell} \left( 1+ \frac{\alpha \Delta}{R^2 s} \right)^{-\Delta_{12,5}} \left( 1 - \frac{\alpha \Delta}{R^2 s} \right)^{-\tilde \Delta_{34,  5}} 
\ea
In the flat space limit, $\Delta_5 \propto R$, so the last two terms can be approximated by exponentials of small quantities.  Note that for $\ell = 0$, the powers of $\alpha$ clearly cancel.  When $\ell > 0$, we must also include the Mack polynomials, which are homogeneous in the flat space limit, providing an overall factor of $\alpha^{-\ell}$ so that again, the power-law dependence on $\alpha$ in the integrand cancels.  We find
\ba
\lim_{R \to \infty} \left( \frac{R^2 s}{2} \right)^{\Delta_{1234} - h- \ell} \int_{- i \infty}^{i \infty} \frac{d \alpha}{2 \pi i} \ \! e^{ \pi i(h- \Delta)} \left( e^{\pi i \frac{R^2s}{2\alpha} } -1 \right) e^{\alpha \left( 1 + \frac{\Delta^2}{R^2 s} \right)}
\ea
But this is just an ordinary fourier transform in $i \alpha$.  The first term in parentheses gives a vanishing contribution for $R \to \infty$, but the second term gives
\be
\left( \frac{M^2}{2} \right)^{\Delta_{1234} - h - \ell}  e^{ \pi i(h- \Delta)}  \delta \left( 1 - \frac{M^2}{ s} \right)
\ee
where $\Delta = RM$.  As expected, we see that the flat space limit of a $4$-pt scalar conformal block is simply a delta function that sets the center of mass energy equal to the mass (dimension) of the conformal block.  

As we show in the appendix, in the case of general spin $\ell$, a conformal block is simply a Mack polynomial $P_{\ell,\Delta}(\delta_{ij})$ multiplied by $B_\Delta^\ell(\delta_{12})$.  The purpose of the Mack polynomial is to encode the angular momentum information, so it is no surprise that in the flat space limit it simply becomes
\be
P_{\ell, \Delta}(\delta_{ij}) \to \left(\frac{M R}{2} \right)^d \left(- \frac{R^2 s}{4 \alpha} \right)^\ell P_\ell^{(d)}(\cos \theta)
\ee
where $P_\ell^{(d)}(\cos \theta)$ are the Legendre or Gegenbauer polynomials appropriate to $d$-dimensional spherical harmonics.  The dependence on $\alpha$ simply compensates for the $\alpha^\ell$ dependence of $B_\Delta^\ell$, as noted above.  So we find that the flat space limit of a general conformal block is
\be
\left[ \left( \frac{M^2}{2} \right)^{\Delta_{1234} }  e^{ \pi i(h- \Delta)} \right] P_\ell^{(d)}(\cos \theta)  \delta \left( 1 - \frac{M^2}{ s} \right)
\ee
as anticipated, up to a convention-dependent normalization factor.

\subsection{Implications of Hawking Evaporation for CFTs}
\label{sec:BH}

Black Hole thermodynamics leads to certain expectations for scattering amplitudes at very high energy and small impact parameter \cite{Giddings:2009gj, Amati:1987wq, Giddings:2011xs}.  In the case of 2-to-2 scattering at trans-Planckian energies, the enormous entropy of macroscopic black holes combined with the tiny entropy of a 2-particle state suggests that the scattering amplitude will be
\be
\mathcal{S}(s) \sim \exp \left[-\frac{1}{2} S_{BH}(s) \right] = \exp \left[- \frac{1}{8} \left( G_{D} s^{\frac{D-2}{2}} \right)^{\frac{1}{D-3}} \right]
\ee
for a $D$ dimensional bulk, so that the cross section is exponentially suppressed by the black hole entropy.  We have included the factor of $\frac{1}{2}$ in the exponent because it is the cross section, not the amplitude, which should be suppressed by the black hole entropy; one should also use the more general Kerr entropy at large impact parameter.   Of course the same result obtains from the thermal spectrum of Hawking radiation.  In principle, the amplitude could be larger than this.\footnote{But it cannot be any smaller, because the probability that the two particles tunnel through each other is of this order. }   However, if two massless particles collide with an impact parameter smaller than the Schwartzchild radius associated with their center of mass energy, then causality suggests that the two particles will be inside a trapped surface before they can interact, and so black hole formation seems unavoidable \cite{Giddings:2011xs}.  

This semiclassical expectation for the high-energy 2-to-2 scattering amplitude leads to a very general prediction for the behavior of the conformal block decomposition in equation (\ref{eqn:ConformalBlockDecomp}) at large dimensions.  In particular, the conformal block coefficients should take the form
\be
C(\Delta, \ell) \sim \exp \left[- \frac{1}{2} \left( \frac{\Delta^{D-2}}{N^2} \right)^{\frac{1}{D-3}} \right] \ \ \ \mathrm{for} \ \ \ \ell \ll \left( \frac{\Delta^{D-2}}{N^2} \right)^{\frac{1}{D-3}} \ \ \ \mathrm{when} \ \ \  \Delta^{D-2} \gg N^2
\ee
where $D$ is the full spacetime dimension of the decompactifying bulk (if there are no compactification manifolds whose size grow with the AdS scale $R$, then $D = d+1$).  We are defining a rough version of the central charge via $N^2 \approx 1/G_{D}$.   This is a prediction for the conformal block decomposition in any CFT with a bulk dual that can be described by a low-energy effective field theory with classical gravity.  

Unfortunately, this whole discussion has been the reverse of what we really desire -- namely  a \emph{derivation} of these results directly from the CFT.  Obtaining such a derivation, which must be a robust consequence in any CFT with a weakly coupled gravity dual, may be the most important problem in quantum gravity.

\section{Discussion}

Locality is one of the most important ideas in physics, and understanding it \cite{JP, Katz, Heemskerk:2010hk, Heemskerk:2010ty, Sundrum:2011ic} has only become more urgent since the discovery that a local spacetime can emerge holographically from a lower dimensional theory, as in the AdS/CFT correspondence \cite{Maldacena, Witten, GKP}.  

However, in the absence of an explicit Lagrangian (and even in its presence \cite{Adams:2006sv}) the definition of locality is often rather murky.    Fortunately, one can make the notion precise in flat spacetime: local theories can be defined as those that give rise to S-Matrices that are analytic and exponentially bounded functions of the kinematic invariants.  In this work we have argued that one can understand the analyticity properties of the S-Matrix in theories of quantum gravity by taking the flat spacetime limit of AdS/CFT correlation functions in the Mellin representation.  We have taken some steps in this direction by showing how the familiar analyticity properties of the S-Matrix in bulk perturbation theory emerge from the meromorphy of the Mellin amplitude, but there is much more work that could be done analyzing the Mellin amplitudes from more interesting CFTs.

When it comes to AdS/CFT, one can discuss bulk locality on scales both larger and smaller than the AdS scale $R$.  On distance scales larger than $R$, bulk locality can be analyzed using the holographic RG \cite{Heemskerk:2010hk, Skenderis:2002wp}, but from this point of view sub-AdS scale locality remains very mysterious.  In regions much smaller than $R$ the AdS curvature should be irrelevant, and so the holographic RG approach must break down, since we do not expect there to be an RG picture when we `integrate out' an approximately flat dimension. Instead, it is much more natural to analyze sub-AdS scale holography using flat space methods, and chief among these is the analyticity of the S-Matrix.  The Mellin amplitude makes this analysis possible for AdS/CFT.

Another motivation for investigating locality is to understand to what extent (if any) black hole evaporation should be viewed as a local process governed by a local theory.  It has been anticipated that new tools are needed to attack this question;  in fact, one of the goals of research into new on-shell methods for describing scattering amplitudes \cite{ArkaniHamed:2008gz, ArkaniHamed:2009dn} has been to give a less manifestly local description of the S-Matrix, in the hopes that such a description would be easier to adapt to the scattering amplitudes of quantum gravity.  We believe that the Mellin amplitude is the appropriate tool for these questions.  Gravitational physics in AdS with $D>3$ and a parametric separation between $R$ and $\ell_{Pl}$ generically involves black holes that are much smaller than the AdS radius.  Thus we expect that the dual CFTs and their Mellin amplitudes will provide an ideal laboratory for investigating Hawking evaporation in a setting where bulk locality is not manifest.  Scattering processes involving black holes will violate the exponential boundedness requirements associated with locality \cite{ArkaniHamed:2007ky, Giddings:2009gj, Amati:1987wq, Giddings:2011xs}, but it will be very interesting to investigate exactly how locality breaks down, and to give a precise criterion for the situations and questions for which effective field theory fails.  

We took the first step towards an analysis of black holes by identifying a robust consequence of Hawking evaporation for the conformal block decomposition.  One expects that 2-to-2 scattering will be exponentially suppressed at energies larger than the Planck scale due to the large black hole entropy, and this implies a corresponding exponential shutdown in the coefficients for the conformal block expansion at large dimensions in the dual CFT.  This is a sharp and generic prediction that could be understood through the direct study of CFTs; such a result would be the first step towards a microscopic derivation of Hawking evaporation.

We have given a non-perturbative, holographic definition of the S-Matrix using the flat spacetime limit of AdS/CFT, so it is natural to ask whether \emph{all} consistent S-Matrices in quantum gravity can be obtained from the large central charge limit of a CFT.  In other words, is AdS/CFT \emph{the} path to quantizing gravity in flat spacetime?  We simply do not know, but it would be fascinating if there exists a generic alternative description.  The remarkably simple properties of the scattering amplitudes themselves \cite{Britto:2004ap, Britto:2005fq, Bern:2006kd, Bern:2009kd, ArkaniHamed:2008yf, Drummond:2008vq, ArkaniHamed:2008gz, ArkaniHamed:2009dn} and the existence of matrix models \cite{Banks:1996vh} might be viewed as evidence for this sort of conjecture.  In any case, our work suggests that there is an intimate connection between the Bootstrap Program for large $N$ CFTs  \cite{Belavin:1984vu, Dolan:2003hv, Rattazzi:2008pe, Costa:2011mg, Costa:2011dw, Hellerman:2009bu, Poland:2011ey, Rychkov:2011et,Fitzpatrick:2011hh}  and the S-Matrix Program, and that solving general CFTs via the Bootstrap will be at least as difficult as finding consistent S-Matrices using the constraints of crossing symmetry, analyticity, and unitarity.  

The investigation of Mellin amplitudes is only its infancy.  In a companion paper \cite{Unitarity} we will show how the usual unitarity relations for the S-Matrix arise from the OPE of the CFT.  However, there are many other elementary questions that still need to be addressed, including the development of diagrammatic rules for loops and for fields with spin.  Here we have only obtained the flat space S-Matrix for external states composed of massless scalar particles, so some further technical developments \cite{Costa:2011mg, Costa:2011dw} will be necessary to treat arbitrary species of particles.  Techniques from the analysis of scattering amplitudes, such as the BCFW recursion relations \cite{Britto:2004ap, Britto:2005fq, ArkaniHamed:2008yf}, can likely also be applied in Mellin space.  It will interesting to see if the Mellin space approach can also offer insight into theories without the full conformal group of symmetries.

\section*{Acknowledgments}

We would like to thank Nima Arkani-Hamed, Daniel Harlow, Ami Katz, Dhritiman Nandan, Miguel Paulos, Matt Reece, Steve Shenker, David Simmons-Duffin, and especially Jo\~ao Penedones for discussions, and Suvrat Raju, Balt van Rees, and Jo\~ao Penedones for early collaboration.  We would also like to thank Edward Witten for inquiries that motivated aspects of appendix \ref{app:DefinitionUniqueness}.   ALF was partially supported by ERC grant BSMOXFORD no. 228169.  JK acknowledges support from the US DOE under contract no.~DE-AC02-76SF00515.

\appendix

\section{Basic Definitions}
\label{app:DefinitionUniqueness}

\subsection{Measure of Integration}
\label{app:MeasureOfIntegration}

Before considering the definition and uniqueness properties of the Mellin amplitude, let us give a precise definition of the measure of integration, as given by \cite{Mack, MackSummary, Penedones:2010ue} .  Given a particular solution $\Delta_{ij}$ to the constraint equations, we can define
\be
\delta_{ij} = \Delta_{ij} + \sum_{k=1}^{\frac{n(n-3)}{2}} c_{ij,k} s_k
\ee
where the coefficients $c_{ij,k}$ are symmetric in $i,j$ vanish when $i=j$, and satisfy
\be
\sum_{j} c_{ij,k} = 0
\ee
so that $\delta_{ij}$ continue to satisfy the constraint equation for arbitrary $s_k$.  Now we can view $c_{ij,k}$ as an square matrix of size $\frac{n(n-3)}{2}$ if we let the $ij$ take on all of their non-zero values.  If we demand that
\be
|\det c_{ij,k}| = 1
\ee
then the measure of integration will simply be
\be
\int [d \delta] = \int_{-i\infty}^{i \infty} \prod_{k=1}^{\frac{n(n-3)}{2}} d s_k
\ee
Note that there are poles from the $\Gamma(\delta_{ij})$ for all non-positive integer values of the $\delta_{ij}$, and also poles (generically, semi-infinite sequences of poles) at positive $\delta_{ij}$ from the Mellin amplitude.  The contour of integration is chosen so that it passes entirely to one side of these semi-infinite lines of poles.  For large values of the dimensions $\Delta_i$ this sort of contour is obtained automatically by giving the $\delta_{ij}$ a small real part, but otherwise integrating along a line from $-i \infty$ to $i \infty$ as indicated.  For example, at 4-pt we can write all of the $\delta_{ij}$ directly in terms of $\delta_{12}$ and $\delta_{13}$, and if all the external dimensions are equal then we find
\be
\int_{\epsilon-i \infty}^{\epsilon+ i \infty} d \delta_{12} d \delta_{13} M(\delta_{ij}) \Gamma(\delta_{12})^2 \Gamma(\delta_{13})^2 \Gamma(\Delta - \delta_{12} - \delta_{13})^2 \prod_{i<j} (x_{ij})^{-2 \delta_{ij}}
\ee
One can immediately see that the factors of $\Gamma(\delta_{12})^2 \Gamma(\delta_{13})^2 $ give poles at non-positive integer values of $\delta_{12}$ and $\delta_{13}$, while the final $\Gamma(\Delta - \delta_{12} - \delta_{13})^2$ factor and the Mellin amplitude itself only have poles at positive values of $\delta_{12} + \delta_{13}$.  The contour of integration naturally runs between these lines of poles, as pictured in figure \ref{fig:PolePrescription}.   The fact that some are double poles signals the presence of anomalous dimensions in perturbation theory (the exact Mellin amplitude will only have single poles).  

\subsection{Mellin Amplitude:  Definition and Non-Uniqueness }

Our goal here is to give a precise definition of the Mellin amplitude, so let's begin with an $n$-pt CFT correlator $A_n(u_{ijkl})$ that has been stripped of some overall dependence on the $x_i$ coordinates, so that it only depends on cross ratios  
\be
u(x_i, x_j, x_k, x_l) = \frac{r_{ij} r_{kl}}{r_{ik} r_{jl}}
\ee  
where $r_{ab} = (x_a - x_b)^2$.   For now we will imagine a putative CFT in infinite $d$, so that we can ignore $d$-dependent relations among the cross ratios.  

Now let us use conformal invariance to set $x_1 = 1, x_2 = 0,x_3 = \infty$ for convenience, and we will freely refer to these as  though they are labeled $x_1, x_0, x_\infty$.  In this frame we find that
\ba
r_{1i} &=& u(0,\infty, 1, x_i) \\
r_{0i} &=& \frac{1}{u(1,0,\infty, x_i)} \\
r_{ij} &=& \frac{u(0,\infty, x_i, x_j)}{u(1,0,\infty, x_i)}
\ea
where in all cases $i, j > 3$.  Since we have that $r_{\infty i} = \infty$ and $r_{01} = 1$, this means that we can express all the non-trivial Poincar\'e invariants $r_{ij}$ in terms of $u(0,\infty, i, j), u(0,\infty, 1, i)$, and $u(1,0,\infty, i)$ and it is easy to count and find that there are $n(n-3)/2$ of them.  So the cross ratios appearing above provide a complete basis of conformal invariants on which $A_n$ can depend.

This accords with a certain basis for the $\delta_{ab}$ of the Mellin amplitude that we wish to define.  Namely, if we use the constraints $\sum_{j} \delta_{ij}  = \Delta_i$ to eliminate the $\delta_{\infty a}$ and $\delta_{10}$ then we are left with exactly $n(n-3)/2$ variables (recall that $\delta_{ij} = \delta_{ji}$ and $\delta_{ii} = 0$).  Thus we can define the Mellin amplitude via the inverse Mellin transform in each variable,
\be \label{InfinitedDefinition}
N(\delta_{1i},\delta_{0i}, \delta_{ij}) = \int_0^\infty A_n(u_{ijkl}) \prod_{i=3}^n (u_{0 \infty 1 i})^{-\Delta_i - \delta_{0i}-1} d u_{0 \infty 1 i}   (u_{1 0 \infty i})^{\delta_{1i}-1} d u_{1 0 \infty i}  \prod_{3<i<j}^n  (u_{0 \infty ij})^{\delta_{ij}-1} d u_{0 \infty ij}  
\ee
We also could have just defined this directly in terms of $r_{0i}$, $r_{1i}$, and $r_{ij}$ for $i,j > 3$.
What we call the Mellin amplitude (as opposed to the integrand $N$ above) can now be defined as
\be
M(\delta_{ab}) = \frac{N(\delta_{1i},\delta_{0i}, \delta_{ij})}{\prod_{a<b} \Gamma(\delta_{ab}) }
\ee
so that
\be
A_n(u_{ijkl}) = \int_{-i \infty}^{i \infty} [d \delta] M_n(\delta_{ab}) \prod_{a<b} \Gamma(\delta_{ab}) r_{ab}^{- \delta_{ab}} 
\ee
as desired.  

Now let us consider the situation when $n > d+2$.  The Mellin amplitude will no longer be uniquely defined, so there will be many Mellin-space functions that give rise to vanishing position-space correlators (these Mellin amplitudes are ``pure gauge'').  However, it is simple to generate these ``pure gauge'' Mellin amplitudes, and to see that they vanish when we take the flat space limit.

Let us universalize the notation by taking the indices $a,b$ to run from $1$ to $d+2$, while the indices $i,j$ run from $d+3$ to $n$, and finally the indices $\alpha, \beta$ can take any value from $1$ to $n$.   It will be useful to represent the kinematics in $d+2$ dimensional embedding space coordinates $P_\alpha^A$ \cite{Weinberg:2010fx}.  These variables are constrained by $P_\alpha^2 = 0$ and identified projectively as $P_\alpha^A \sim \lambda P_\alpha^A$ for positive real $\lambda$.  We can choose an explicit $P_\alpha^A$ corresponding to $x_\alpha^\mu$ by taking $(P_\alpha^-, P_\alpha^+, P_\alpha^\mu) = (1,x_\alpha^2, x_\alpha^\mu)$, so that $(x_\alpha - x_\beta)^2 = P_\alpha \cdot P_\beta$ and we will use the shorthand $P_\alpha \cdot P_\beta = P_{\alpha \beta}$.

Now let us define the $d+3$ by $d+3$ matrix
\be
A^{\alpha \beta}(I,J) = \left[ \begin{array}{cc}
P_{ab} & P_{aJ}  \\
P_{Ib} & P_{IJ} \end{array} \right]
\ee
where the indices $I$ and $J$ are fixed, whereas the $a$ and $b$ indices range from $1$ to $d+2$ (so $P_{ab}$ forms a $d+2$-by-$d+2$ sub-matrix in the upper left corner, etc).  

The point is that $\det[A(I,J)]  = 0$ for all $I$ and $J$ between $d+3$ and $n$.  This follows immediately from the fact that the entries of $M(I,J)$ are formed from the matrix product $P_\alpha^A P_{A \beta}$, and when we regard the object $P_\alpha^A$ as a $d+2$ by $n$ matrix it has rank at most $d+2$.  This means that generically there is an explicit formula relating
\be
P_{IJ} = f_{IJ}(P_{ab}, P_{aI}, P_{aJ})
\ee
by using this determinant condition.  This reduces the full space of $P_{\alpha \beta}$ to a 
\be
\frac{(d+2)(d-1)}{2} + d(n-d-2) = \frac{d(2n-d-3)}{2} - 1
\ee 
dimensional space (when $n \geq d+2$, otherwise the result is just $n(n-3)/2$), because the unconstrained first $d+2$ points give $(d+2)(d-1)/2$ degrees of freedom and then each extra point beyond $d+2$ gives $d$ extra degrees of freedom.  

It is tempting to try to use this reduced set of variables to define the Mellin amplitude, or at least to show that it exists, by taking the Mellin transform in only this reduced set of variables.  However, this would break the permutation symmetry among the external operators in an ad hoc way, and it is unclear how it would be related to our known results for Mellin amplitudes corresponding to tree-level Witten diagrams, which do not require a choice of preferred basis or ``gauge fixing''.  

However, another way of looking at the situation is that while the fact that $\det[A(I,J)] = 0$ is fairly clear in position space, it becomes a non-trivial statement in Mellin space, where the spacetime dimension makes no explicit appearance.  Thus if we take a conformally covariant function, express it as an integral over Mellin space, multiply by $\det[A(I,J)]$, and interpret the result as a single  Mellin space amplitude then we obtain an equality that identifies functions on Mellin space that integrate to zero.  In other words, we have a way of finding functions in Mellin space that are ``pure gauge''.  Applying this procedure shows that for any Mellin integrand function $N(\delta_{\alpha \beta})$, meaning any function where the integral 
\be
A(P_\alpha) = \int [d \delta] N(\delta_{\alpha \beta}) \prod_{\alpha < \beta}^n (P_{\alpha \beta})^{- \delta_{\alpha \beta}}
\ee
gives a well-defined result, we have (in $d$ dimensions)
\be
0 = \int [d \delta] \prod_{\alpha < \beta}^n (P_{\alpha \beta})^{- \delta_{\alpha \beta}}   \left[ \sum_{\sigma(a_1,...,a_n)} \mathrm{sign} (\sigma) N \left(\delta_{1 s_{\sigma(1)}} + 1, \delta_{2 s_{\sigma(2)}} + 1 ..., \delta_{m s_{\sigma(m)}} + 1 \right) \right]
\ee
where we are summing over all permutations $\sigma$ on $m = d+3$ or more indices, and the $s_i$ are $m$ arbitrary distinct integer labels ranging from $1$ to $n$.  The $\delta_{\alpha \beta}$ that have not been explicitly written are unchanged.   This formula is simply implementing the determinant --  the shifts in the Mellin variables occur because we have soaked up each monomial in the determinant by a shift of the $\delta_{\alpha \beta}$ variables.  There will also be even more general objects that do not use the labels $1,2,...,m$ but any set of distinct integers, but we have written the above as a simpler representative example. 

The object in square brackets is ``pure gauge'' as a Mellin integrand.  This object will also vanish when we take the flat space limit to compute the S-Matrix.  The reason is that when we perform this procedure, the Mellin amplitude (as opposed to the integrand $N$ above) will inherit an overall factor of 
\be
\det \left[ \begin{array}{cc}
\delta_{ab} & \delta_{aJ}  \\
\delta_{Ib} & \delta_{IJ} \end{array} \right]
\ee
due to the shifts $\delta_{i s_{\sigma(i)}} \to \delta_{i s_{\sigma(i)}} + 1$ applied to the $\Gamma(\delta_{i s_{\sigma(i)}})$ functions.  But in the flat space limit this is just
\be
\det \left[ \begin{array}{cc}
p_a \cdot p_b & p_a \cdot p_J  \\
p_b \cdot p_I & p_I \cdot p_J \end{array} \right]
\ee
which vanishes in $d+1$ dimensional momentum space for the same reason that $\det[M(I,J)]$ vanished in position space in the CFT.  Thus the flat space limit of ``pure gauge'' Mellin amplitudes will be zero.

\section{Derivation of the Flat Space Limit Formula}

\subsection{Single Particle Normalizations}
\label{sec:SingleParticleNorms}

In the normalization conventions from \cite{Fitzpatrick:2011ia}, the two-point function of $\CO$ is 
\ba 
\< \CO( t_1, \hat{x}_t) \CO(t_2, \hat{x}_2 ) \> &=& \CC_\Delta \sum_{n,\ell} \frac{e^{i \omega_{n,\ell} (t_1 - t_2)}}{N_{n \ell J}^2} Y_{\ell J}(\hat{x}_1) Y^*_{\ell J}(\hat{x}_2),
\ea
So, taking the smearing of the operators, we have
\ba
\< \omega, \hat{v} | \omega', \hat{v}' \> &=&  \int_{-\tau}^\tau dt_1 dt_2 e^{i (\omega_1 t_1 - \omega_2 t_2)} \CC_\Delta \sum_{n, \ell} \frac{e^{i \omega_{n, \ell} (t_1 - t_2)} }{N_{n,\ell, J}^2} Y_{\ell}( \hat{x}_1) Y_{\ell} (\hat{x}_2) \nn\\
 &=&
 \sum_{n,\ell} (2 \pi)^2 \delta_\tau (\omega_1-\omega_{n,\ell}) \delta_\tau (\omega_2 - \omega_{n,\ell}) \left( \CC_\Delta \frac{2 \pi^h}{\Gamma(\Delta)} \right)^2 \left( \frac{R\omega_{n,\ell}}{2} \right)^{2\Delta - 2h} Y_\ell(\hat{x}_1) Y^*_\ell(\hat{x}_2) \nn\\
 &=& 2 \pi^2 R \delta_\tau (\omega_1 - \omega_2) \left( \CC_\Delta \frac{2 \pi^h}{\Gamma(\Delta)} \right)^2 \left( \frac{R \omega_1}{2} \right)^{2\Delta-2h} \delta(\hat{v}_1, \hat{v}_2) \nn\\
 &=&  R^{2\Delta -2h+1} \left[ 2 \omega \delta( \vec{p}_1 - \vec{p}_2) \right] \left[ \frac{\pi^2}{2^{2\Delta-2h}} \left( \omega^{\Delta-1} \frac{\CC_\Delta 2 \pi^h}{\Gamma(\Delta)} \right)^2 \right],
\ea
where we have used $\delta^{(d)}(\vec{p}- \vec{p}') \sim \frac{1}{p^{d-1}} \delta(\hat{p}, \hat{p}') \delta(|p|-|p'|)$ and $\frac{1}{N_{n,\ell,J}^2} \sim 
n^{2\Delta-2h} \CC_\Delta \left( \frac{2 \pi^h}{\Gamma(\Delta)} \right)^2 $.  This provides the normalizations in equation (\ref{eq:SingleParticleNorm}).

\subsection{Completing the Derivation}
\label{app:GaussianIntegrals}

Our starting point is the simple formula we left off from at the end of section \ref{sec:FlatSpaceLimit}, allowing a non-zero energy violation component $q_0$:
\be
\exp \left[ i t_\omega + \frac{t_\omega^2}{4\alpha}- \frac{n q_0 \sum_i t_i^2 \omega_i}{4 \alpha}  - \frac{R^2}{4\alpha}\left( -(nq)^2 + \sum_{i<j} \frac{u_{ij}^2}{2 s'_{ij}} \right) \right]
\ee
where we have defined $t_\omega = \sum_i t_i \omega_i $ and
\ba
u_{ij} &=&  \epsilon_{ij} - \frac{2n}{n-2} q \cdot (p_i + p_j) + \frac{t_{ij}^2 \omega_i \omega_j}{R^2} .
\ea
Now we need to integrate over the constrained $u_{ij}$ and $t_i$ variables and multiply by the single-particle normalizations to obtain our desired result, equation (\ref{FlatSpaceLimitFormula}).  We will implement the $n+1$ constraints on the $u_{ij}$ variables 
\ba
0 &=& \sum_{j\ne i} u_{ij} + \frac{2n}{n-2} q \cdot (p_i + p_j) - \frac{t_{ij}^2 \omega_i \omega_j}{R^2}   \\
0 &=& u_{12} + \frac{2n}{n-2} q \cdot (p_1 + p_2) - \frac{t_{12}^2 \omega_1 \omega_2}{R^2}  
\ea
via Lagrange multipliers.  We have made an arbitrary choice to isolate $u_{12}$, which is necessary so that $\epsilon_{ij}$ do not include the $\alpha$ direction of integration; this means that there will be a factor of $s_{12}$ from the change of variables from $\delta_{ij}$ to $\{  \alpha$, $u_{ij} \}$.  Thus we have the exponential integrand
\ba
&& \exp \left[ i \Sigma_i t_i \omega_i + \frac{(\Sigma_i t_i \omega_i)^2}{4\alpha}  - \frac{n q_0 \sum_i t_i^2 \omega_i}{4 \alpha} - \frac{R^2}{4\alpha}\left( -(nq)^2 + \sum_{i<j} \frac{u_{ij}^2}{2 s'_{ij}} \right) \right.  \\ 
&& \left. + i \sum_i \lambda_i \left( \sum_{j\ne i} u_{ij} + \frac{2n}{n-2} q \cdot (p_i + p_j) - \frac{t_{ij}^2 \omega_i \omega_j}{R^2}   \right)
 +i\lambda \left( u_{12} + \frac{2n}{n-2} q \cdot (p_1 + p_2) - \frac{t_{12}^2 \omega_1 \omega_2}{R^2}  \right)
  \right] \nn
\ea
with Lagrange multipliers $\lambda_i$ and $\lambda$.  Now we can integrate over the $u_{ij}$ without any constraints.  This will give an overall factor of 
\be
\left( \frac{4 \alpha}{R^2} \right)^{\frac{n(n-1)}{4}} \prod_{i<j}^n \sqrt{s_{ij}'}
\label{eq:sijprefactor}
\ee
nicely canceling many terms from the prefactor in equation (\ref{eq:Prefactor}).  In the exponent we have (after simplifying and using the constraint that $\sum_{j\neq i} s_{ij}' = 0$):
\ba
&& \exp \left[ i \Sigma_i t_i \omega_i + \frac{(\Sigma_i t_i \omega_i)^2}{4\alpha}  + \frac{R^2}{4\alpha} (nq)^2 - \frac{n q_0 \sum_i t_i^2 \omega_i}{4 \alpha}  + \frac{2 \alpha}{R^2} S_{ab} \lambda_a \lambda_b \right.   \\ 
&& \left. + i \sum_i \lambda_i \left( \sum_{j\ne i} \frac{2n}{n-2} q \cdot (p_i + p_j) - \frac{t_{ij}^2 \omega_i \omega_j}{R^2}   \right)
 +i \lambda \left( \frac{2n}{n-2} q \cdot (p_1 + p_2) - \frac{t_{12}^2 \omega_1 \omega_2}{R^2}  \right)
  \right] \nn
\ea
where $a$ runs from $0$ to $n$, with $\lambda_0\equiv \lambda$, and $S_{ab}$ is
\ba
  && S_{ab} = \left( \begin{array}{cccccc} s'_{12} & s'_{12} & s'_{12} & 0 & 0 & \dots \\
  s'_{12} & 0 & s'_{12} & s'_{13} & s'_{14} & \dots \\
  s'_{12} & s'_{12} & 0 & s'_{23} & s'_{24} & \dots \\
  0 & s'_{13} & s'_{23} & 0 &  s'_{34} & \dots \\
0 & s'_{14} & s'_{24} & s'_{34} & 0 & \dots \\
  \vdots & \vdots & \vdots & \vdots & \vdots & \ddots \end{array} \right), 
  \ea
so now we can integrate over the Lagrange multipliers.  This produces a prefactor that is parametrically of the form
\be
\left( \frac{R^2}{2 \alpha} \right)^{\frac{n+1}{2}} \frac{1}{ \sqrt{\det[S ]}}
\label{eq:Sprefactor}
\ee
and returns an exponential
\ba
&& \exp \left[ i \Sigma_i t_i \omega_i + \frac{(\Sigma_i t_i \omega_i)^2}{4\alpha}  + \frac{R^2}{4\alpha} (nq)^2 - \frac{n q_0 \sum_i t_i^2 \omega_i}{4 \alpha}   + \frac{R^2}{8 \alpha} \sum_{a, b} U_a U_b  [S^{-1}]_{ab} 
  \right] \nn \\
 && U_0 = v_{12} - \frac{t_{12}^2 \omega_1 \omega_2}{R^2}, \ \ \ \ \ U_i = \sum_{j\ne i} v_{ij} - \frac{t_{ij}^2 \omega_i \omega_j}{R^2},
\ea
where for notational convenience we have defined the quantity $v_{ij} = \frac{2n}{n-2} q \cdot (p_i + p_j)$, and it will also be useful to define $v_a = \sum_{j\ne i} v_{aj} = 2n q \cdot p_a + \frac{2n^2}{n-2} q^2 \approx 2n q \cdot p_a$.   Note that here we have assumed that we have generic momenta with det$[S] \neq 0$ and $d+3 > n$.  One can check that in the case $n > d+2$ the same final result obtains.

Finally, we must integrate over the times $t_i$, ignoring the negligible quartic piece which is suppressed by additional powers of the AdS length $R$.  We must also include a regulator associated with the temporal boundaries at $t_i = \pm \tau$; to simplify the integrals we will replace this hard boundary with Gaussian factors $\exp \left[-\sum_i \frac{ t_i^2}{2\tau^2} \right]$ (it is important to maintain non-overlapping support between the in and out states, as emphasized in \cite{Gary:2009mi}, but this will not an issue at this point in the derivation, where we need only evaluate some Gaussian integrals in the simplest way available).  
Now we are left with time integrals over the integrand
\ba
 &&\equiv \exp \left[  \frac{R^2}{4\alpha} (nq)^2 + \frac{R^2}{8 \alpha} v_{12}(v_{12} + 4 n q \cdot \sum_i p_i (S^{-1})_{0i} )
 + \frac{R^2}{2\alpha} \sum_{ij} n^2 (q \cdot p_i ) ( q \cdot p_j) (S^{-1})_{ij} \right. \nn\\
 &&\ \ \ \ \ \ \ \ \left. + i \Sigma_i t_i \omega_i-\frac{1}{2} \sum_{ij} t_i t_j A_{ij}
 \right]  
 \label{eq:timeintegrals}
 \ea
where $A_{ij}$ takes the form
 \ba
 A_{ij} &=&- \delta_{ij} \frac{1}{2 \tau^2} + \frac{\omega_i \omega_j}{4 \alpha} + Q_{ij}
 \ea
 where $Q_{ij} $ is $\CO(q)$ and higher, so its inverse is
 \ba
 A^{-1}_{ij} &\approx& Y (1-QY + (QY)^2)_{ij} + \CO(q^3)  \nn\\
 Y_{ij} &=& -2 \tau^2 \delta_{ij} + \omega_i \omega_j \left(  \frac{2 \tau^2}{\sum_k \omega_k^2} + \frac{4 \alpha }{(\sum_k \omega_k^2)^2} \right) + \CO(\frac{1}{\tau})
 \ea
 Note that $Y_{ij} \omega_j = \left(  \frac{4 \alpha}{\sum_k \omega_k^2} \right) \omega_i$, so $\frac{1}{4}\omega_i A^{-1}_{ij} \omega_j$ simplifies
 to
 \ba
 \frac{1}{4} \omega_i A^{-1}_{ij} \omega_j &=&  \alpha - \left( \frac{2 \alpha }{\sum_k \omega_k^2} \right)^2 \omega_i (Q - QYQ)_{ij} \omega_j
  + \CO(q^3)
  \label{eq:timegauss}
  \ea
The matrix $Q$ has four pieces, from the exponential above:
\ba
Q &=& Q^{(1)} + Q^{(2)} + Q^{(3)} +Q^{(4)}\nn\\
Q^{(1)}&=&  \frac{\omega_1 \omega_2 }{2\alpha} (v_{12} S^{-1}_{00} + 2 n q \cdot \sum_i p_i S^{-1}_{0i} )
 \left( \begin{array}{ccc c}
 1 & -1 & & \\ 
 -1& 1 & & \\
  & & 0 \\
   & & & \ddots \end{array} \right)
    \nn\\
   Q^{(2)}_{ij} &=& \frac{ v_{12}}{\alpha} \left\{ \begin{array}{ll} \omega_i \omega_j (S^{-1}_{0i}+ S^{-1}_{0j}) & i \ne j \\
    - \sum_{k\ne i} \omega_i \omega_k (S^{-1}_{0i} + S^{-1}_{0k}) & i = j \end{array} \right\} \nn\\
     Q^{(3)}_{ij} &=&  \frac{1}{\alpha} \sum_k \left\{ \begin{array}{ll} -2n \omega_i \omega_j q\cdot p_k (S^{-1}_{ik}+S^{-1}_{jk}) & i \ne j \\
  2 n \sum_{\ell \ne i}  \omega_i \omega_\ell q \cdot p_k (S^{-1}_{ik} + S^{-1}_{\ell k}) & i = j
     \end{array} \right\}  \nn\\
     Q^{(4)}_{ij} &=& -\delta_{ij} \frac{n q_0 \omega_i}{4 \alpha} 
\ea
Finally, we want to read off the coefficient of the momentum-conserving $\delta$-function.  Several components of $q$ get large, $\sim \tau^2 q^2$ contributions in the exponent, from the $Q^2$ term in (\ref{eq:timegauss}).  We can see already that there will generally be exactly $n-1$ such $q^\mu$ components (one time and $n-2$ spatial), for the simple reason that $\vec{q}$ always enters in $Q$ dotted into $p_i$ directions.  Naively, this lifts $n$ components of $\vec{q}$; however, not all the $p_i$ are linearly independent.  At leading order, the $p_i$'s conserve momentum, which imposes one constraint.  Furthermore, the symmetries of AdS in the flat-space limit become the symmetries of flat space, and boosts remove an additional independent component of the $p_i$'s.  This is especially obvious in practice, where one often chooses the particles in the ``in'' state to be in their rest frame, so there is already one linear relation among just the incoming momenta.  Thus, at most $n-2$ $\vec{q}$ components can get lifted by $\tau^2$, and for generic external momentum, exactly this many get lifted.  

Focusing on terms $\sim \tau^2 q^2$ in (\ref{eq:timegauss}), we see that they have the form
\ba
 \left( \frac{2 \alpha}{\sum_k \omega_k^2} \right)^2  \omega_i (Q YQ)_{ij} \omega_j &=&
    -2\tau^2 \left( \frac{2 }{\sum_k \omega_k^2} \right)^2 Q_\omega^i \Pi^{ij}_\omega Q_\omega^j + \CO(\tau^0)
    \nn \\ 
    Q_\omega^i \equiv -\frac{1}{2\alpha}  \sum_j Q_{ij} \omega_j, && \  \Pi^{ij}_\omega \equiv \delta_{ij} - \frac{\omega_i \omega_j}{\sum_k \omega_k^2} .
    \ea

Now,
\ba
Q_\omega^i &=& Q^{(1)i}_\omega + Q^{(2)i}_\omega + Q^{(3)i}_\omega +Q^{(4)i}_\omega, \nn\\
Q^{(1)i}_\omega &=& q\cdot \left( -2 \omega_1 \omega_2(\frac{n}{n-2}(p_1 + p_2) S_{00}^{-1} + 2 n \sum_i p_i S_{0i}^{-1} ) (\omega_1 - \omega_2) \right)
(\delta_{i1} - \delta_{i2})
 \nn\\
Q^{(2)i}_\omega &=& q\cdot \left( -\frac{4n}{n-2} (p_1 + p_2) \right)  \sum_j \omega_i \omega_j (\omega_j - \omega_i) (S_{0i}^{-1} + S_{0j}^{-1}) , \nn\\
Q^{(3)i}_\omega &=& q\cdot \left( -2 \sum_{j,k} (-2n) \omega_i \omega_j (\omega_j - \omega_i)p_k (S_{ik}^{-1} + S_{jk}^{-1}) \right) \nn\\
Q^{(4)i}_\omega &=& \frac{n q_0 \omega_i^2}{8}
\ea
Without loss of generality, we can consider our amplitude to be 2 to $n-2$ and we can take particles 1 and 2 in their
rest frame, so that $Q^{(1)}=0$ and $(p_1 + p_2)^\mu = 2\omega_1 \delta^{\mu 0}$.  Since the transition amplitude is proportional to an energy-momentum-conserving $\delta$ function in $p_{\rm tot}^\mu = n q^\mu$, we want to extract the usual scattering amplitude by integrating over $d^{d+1} p_{\rm tot}$.  The time integrations produce a factor of
\be
\sim \sqrt{\det{}' \ Y} \propto \tau^{n-1} \left( \frac{\alpha}{\sum_k \omega_k^2 } \right)^{1/2} .
\label{eq:Yprefactor}
\ee
Here, $\det'$ indicates that all vanishing eigenvalues are discarded.  The $p_{\rm tot}$ integrations produce a factor of 
\be
\sim \left[ \frac{\sum_k \omega_k^2}{\tau} \right]^{n-1} 
\left[ \frac{1}{n^2} \det{}' \frac{\partial}{\partial q^\mu} \frac{\partial }{\partial q^\nu} Q_\omega \cdot \Pi_\omega \cdot Q_\omega \right]^{-\frac{1}{2}} \left[ \frac{\alpha}{R^2} \right]^{\frac{d+1-(n-1)}{2}} 
\label{eq:Qprefactor}
\ee
where the first two terms in brackets come from the $\CO(\tau^2 q^2)$ pieces in (\ref{eq:timegauss}).  The last term in brackets comes from the fact that after $n-1$ $q^\mu$ directions get lifted by $\tau^2$, there are still $d+1-(n-1)$ $q^\mu$ directions that are lifted only by the 
$-\frac{R^2}{2 \alpha} (n q)^2$ term in eq. (\ref{eq:timeintegrals}).  Notice that this exactly cancels the factors of $\tau$ from the time integrations.
Putting together eqs. (\ref{eq:SingleParticleNorm}), (\ref{eq:Prefactor}),(\ref{eq:sijprefactor}),(\ref{eq:Sprefactor}),(\ref{eq:Yprefactor}), and (\ref{eq:Qprefactor}), as well as an additional factor of $s_{12}$ that should be included in the change of variables from $\delta_{ij}$ to $\alpha, \epsilon_{ij}$ due to our choice of gauge $\epsilon_{12}=0$, the total prefactor for the flat-space formula integrand is then (neglecting constants)
\be
\NN^{-1}   \alpha^{h-\Delta_\Sigma}   \left[ s_{12}\left( (\det S) (\det{}' \frac{\partial}{\partial q^\mu} \frac{\partial}{\partial q^\nu} Q_\omega \cdot \Pi_\omega \cdot Q_\omega) \right)^{-\frac{1}{2}}\prod_{i=1}^n \omega_i \left( \sum_k \omega_k^2 \right)^{n-\frac{3}{2}}  \right]
\ee 
Note that the term in brackets is dimensionless, and so can have dependence only on the scattering angles.  Although it is not manifest from the above expression, we have checked for $n=4$ that the dependence on angles cancels. We can easily prove that such a cancellation must occur in all cases, as follows.  The dependence on the angles is independent of the specific theory, and is simply a prefactor that depends only on the momenta of the external particles.  Thus, this prefactor is completely fixed by a single example.  Since the Mellin amplitude is just a constant for $g \phi^n$ theory, and the flat-space S-matrix is as well, this shows that the term in brackets is independent of angles in all cases, thus finishing the derivation of the flat-space S-matrix formula (\ref{eqn:PenedonesConjectureIntro}).

\section{K\"allen-Lehmann in AdS}
\label{app:KLDerivation}

The key to being able to compute a large class of loops is the fact that
\be
\prod_i G_{\Delta_i} (X,Y) = \sum_\alpha N_\alpha G_{\Delta_\alpha}(X,Y)
\ee
for bulk-to-bulk propagators.  One can easily compute $N_\alpha$ for any number of propagators using this simple relation between products and sums of propagators
\ba
G_{\Delta_1}(X,Y) G_{\Delta_2}(X,Y) &=& \sum_n  a_{\Delta_1, \Delta_2}(n) G_{\Delta_1 + \Delta_2+2n}(X,Y), \nn\\
a_{\Delta_1,\Delta_2}(n) &=&  \frac{(h)_n}{2 \pi^h n! }  \frac{  (\Delta_1 + \Delta_2 + 2n)_{1-h}  (\Delta_1 + \Delta_2 + n -2h +1)_n }
{(\Delta_1 + n)_{1-h} (\Delta_2 + n)_{1-h} (\Delta_1 + \Delta_2 + n -h)_n }.
\ea
which was equation (\ref{eqn:ProductOfPropagators}) in the text.  The bulk-to-bulk propagators have a simple expression in terms of the geodesic distance $\sigma$ between $X$ and $Y$:
\ba
G_\Delta (X,Y) &=& \CC_\Delta z^{\Delta/2} {}_2F_1(\Delta, h , \Delta+1-h,z) ,
\ea
where $z=e^{-2\sigma}$.\footnote{This normalization of the bulk-to-bulk propagator is chosen to agree with the two-point functions in \cite{Penedones:2010ue, Fitzpatrick:2011ia}, and differs from that in \cite{Katz} by a factor of $\CC_\Delta$.  Other useful representations of the bulk-to-bulk propagator are
\ba
G_\Delta(X,Y) &=& \CC_\Delta \frac{y^{\Delta/2}}{2^\Delta} {}_2F_1(\frac{\Delta}{2}, \frac{\Delta}{2}+\frac{1}{2} , \Delta+1-h, y) \\
  &=& \frac{\CC_\Delta}{u^\Delta} {}_2F_1(\Delta, \Delta-h + \frac{1}{2} , 2 \Delta -2h +1, -\frac{4}{u}),
\ea
where $u = (X-Y)^2$ and $y^{-\frac{1}{2}} = \cosh \frac{\sigma}{R} = 1 + \frac{u}{2R^2}$.  
} This formula in the case $\Delta_1 = \Delta_2$ was derived in section 3.2 of \cite{Katz} using an inner product on the space of propagators
\be
\left\langle G_{2h - \alpha}, G_\beta \right\rangle = \oint \frac{dz}{2 \pi i} \frac{(1-z)^{2h}}{z^{1+ h}} G_{2h - \alpha}(z) G_\beta(z) = \CC_\alpha \CC_{2h-\alpha} \delta_{\alpha \beta}
\ee
where as usual $d = 2h$ is the dimension of the CFT, $\delta_{\alpha \beta}$ is a kronecker delta, and $z=e^{-2\sigma(X,Y)}$ where $\sigma(X,Y)$ is the geodesic distance between $X$ and $Y$ in the bulk of AdS. 
The generalization that we have used can be derived by computing the inner product
\be
\left\langle G_{\Delta_1 + \Delta_2 + 2n}, G_{\Delta_1} G_{\Delta_2} \right\rangle = a_{\Delta_1, \Delta_2}(n) 
\ee
which vanishes unless $n$ is a non-negative integer.

\section{Conformal Blocks in Mellin Space}
\label{sec:ConformalBlocks}

Up to one very simple but important detail, the conformal blocks corresponding to the exchange of operators of arbitrary spin between  external scalars were constructed by Mack \cite{Mack}.  

First the results.  Let us define the scalar block function
\be \label{eqn:BlockFunction}
B_\Delta^\ell(\delta) = e^{ \pi i(h- \Delta)} \left( e^{i \pi (\delta+ \Delta -2h)} -1\right)  \frac{\Gamma \left(\frac{\Delta-\ell-\delta}{2} \right) \Gamma \left(\frac{2h-\Delta - \ell - \delta}{2} \right) }{ \Gamma \left(\Delta_a-\frac{\delta}{2} \right) \
 \Gamma \left(\Delta_b - \frac{\delta}{2} \right) } 
\ee
where as in the text we have defined the variables $\delta =  \Delta_1 + \Delta_2  - 2 \delta_{12}$, $2\Delta_a = \Delta_1 + \Delta_2$ while $2\Delta_b = \Delta_3 + \Delta_4$ are defined for convenience.  In the scalar case ($\ell = 0$) this function is the conformal block.  The only factor in $B_\Delta^\ell(\delta)$ that Mack \cite{Mack} did not include is the pre-factor in parentheses, which cancels the `shadow' poles from the second $\Gamma$ function in the numerator, so that $B_\Delta^\ell(\delta)$ only has physical poles corresponding to a primary operator and its descendants.

In the case of non-zero $\ell$ we must multiply $B_\Delta^\ell(\delta)$ with $\tau = \Delta - \ell$ by a Mack polynomial \cite{Mack} 
\be \label{eqn:MackPoly}
P_{\ell, \tau}(\delta_{ij}) = \sum_{k=0}^{[\ell/2]} a_\ell^d(k) \left(\delta'_{12} - \frac{\ell}{2} \right)_k \left(\delta'_{34} - \frac{\ell}{2} \right)_k Q_{\ell - 2k}(\delta_{ij})
\ee
where the primed $\delta_{ij}$'s are shifted:
\be
\delta_{12}' = \delta_{12} +\frac{ -\Delta_1 - \Delta_2+d - \Delta}{2}, \ \ \ \ \ \delta'_{34}=\delta_{34}+ \frac{-\Delta_3 - \Delta_4 + \Delta}{2} ,
\ee 
and the $a_\ell^d(k)$ are the coefficients of the familiar Legendre polynomials $P^{(d)}_\ell(\cos \theta)$, which in general dimension are Gegenbauer polynomials $C_\ell^{(h-1)}(\cos \theta)$,  as polynomials in $\cos \theta$:
\ba
\frac{\ell!}{(h-1)_\ell} C_{\ell}^{(h-1)} (t) &=& \sum_k a^d_{\ell, k} t^{\ell - 2k}, \ \ \ \ \ \ a^d_{k,\ell} = \frac{(-1)^k \ell! (h+\ell-1)_k 2^{\ell-2k}}{k!(\ell-2k)!}.
\ea
Mack defines (in our notation) 
\ba \label{eqn:QPoly}
Q_m(\delta_{ij}) &=& 2^{-m} m! \sum_{\sum' k_{ij} = m} (-1)^{k_{14} + k_{23}} \prod' \frac{(\delta_{ij})_{k_{ij}}}{ k_{ij}!} \\
&&
\times \left[ \Gamma\left( \frac{\Delta+\Delta_{12}  - m}{2} + k_{13}+k_{14} \right)
\Gamma\left(\frac{\Delta-\Delta_{12} - m }{2} + k_{23}+ k_{24} \right)\right]^{-1} \nn\\
&& \times \left[
\Gamma\left(\frac{2h-\Delta+\Delta_{34} -m}{2} +k_{13} + k_{23} \right)
\Gamma\left(\frac{2h-\Delta-\Delta_{34} -m}{2} + k_{14}+k_{24}\right) \right]^{-1}, \nn
\ea
where $\Delta_{ij} = \Delta_i - \Delta_j$.  
The notation needs some explanation.  The variables $k$ only connect $1,2$ to $3,4$, so in other words we only have $k_{13}, k_{14}, k_{23},$ and $k_{24}$, and $k_{ij} = k_{ji}$.  So the $\prod'$ terms are  products over the non-vanishing $k_{ij}$.

In the text we discussed the flat space limit of $B_\Delta^\ell(\delta)$.  The Mack polynomials simply reduce to Legendre polynomials in the flat space limit, as should be expected, because their only purpose is to encode the angular momentum information in the conformal blocks.  Let us see why this follows at a technical level.  First, the $\Gamma$ functions in equation (\ref{eqn:QPoly}) cancel when $\Delta \to \infty$ to give an overall factor of $(\Delta/2)^d$.   Now we can perform the sum over the $k_{ij}$.  In the flat space limit, $\delta_{13}, \delta_{24} \propto t$ and $\delta_{14}, \delta_{23} \propto u$.  This means that we can combine terms to find that 
\ba
Q_m(\delta_{ij}) &\to& 2^{-m}  \left(\frac{\Delta}{2} \right)^{-d} \sum_{k=0}^m (-1)^k \frac{m!}{k! (m-k)!} t^k u^{m-k} \nn \\ 
 &=&  \left(\frac{\Delta}{2} \right)^{-d}  \left( \delta_{12} \cos \theta \right)^m
\ea
as desired, where we used the fact that $(t-u)/s = \cos \theta$ in the center of mass frame.  Plugging this result back into (\ref{eqn:MackPoly}) we find that the Mack polynomials reduce 
\be
P_{\ell, \tau}(\delta_{ij}) \to \left(\frac{M R}{2} \right)^{-d} \left( -\frac{R^2 s}{4 \alpha} \right)^\ell \sum_{k=0}^{[\ell/2]} a_\ell^d(k) \cos^{\ell - 2k} \theta
\ee
where we have taken $\delta_{ij} \to -\frac{R^2 s_{ij}}{4\alpha}$ as is appropriate for the flat spacetime limit, and we have assumed that $\Delta = M R$ so that this is a conformal block corresponding to a non-zero energy in the flat space limit.  

The function $B_\Delta^\ell(\delta)$ can be most easily understood as a simple solution to the functional equation we found (\ref{eqn:FunctionalEquation}).  Recall that this equation is simply the eigen-equation of the conformal casimir, so its solutions will be, by definition, the conformal partial waves.  The conformal blocks are simply the conformal partial waves with the correct boundary conditions.  Anyway, in this case the functional equation takes the very simple form 
\be
\frac{(\delta - \Delta+\ell)(2h - \Delta - \ell - \delta)}{(\Delta_a -\delta) (\Delta_b - \delta)} B_\Delta^\ell(\delta) =  B_\Delta(\delta -2) .
\ee
and it is obvious that the two factors in the numerator and denominator are exactly reproduced by each of the four $\Gamma$ functions in equation (\ref{eqn:BlockFunction}).  Note that since this is a finite difference equation, we are free to multiply by any periodic function with period $2$.  The function in parentheses is the choice which eliminates the unphysical shadow poles and gives the correct boundary conditions for the conformal block in position space. 

\bibliographystyle{utphys}
\bibliography{Mellinbib}

 \end{document}